%% file: main.tex
\documentclass[acmsmall,authorversion,nonacm]{acmart}
\setcopyright{none}
\usepackage{booktabs}   
\usepackage{subcaption} 
\usepackage{listings}
\usepackage{todonotes}
\usepackage{physics}
\usepackage{bm}
\usepackage{wasysym}

\begin{document}

\title{Compiling Structured Tensor Algebra}

\author{Mahdi Ghorbani}
\email{mahdi.ghorbani@ed.ac.uk}
\affiliation{
  \institution{University of Edinburgh}            
  \country{United Kingdom}                    
}
\author{Mathieu Huot}
\email{mathieu.huot@stx.ox.ac.uk}
\affiliation{
  \institution{University of Oxford}            
  \country{United Kingdom}                    
}
\author{Shideh Hashemian}
\email{shideh.hashemian@aut.ac.ir}
\affiliation{
  \institution{Amirkabir University of Technology}            
  \country{Iran}                    
}
\author{Amir Shaikhha}
\email{amir.shaikhha@ed.ac.uk}
\affiliation{
  \institution{University of Edinburgh}            
  \country{United Kingdom}                    
}

\renewcommand{\shortauthors}{Ghorbani et al.}

\input{macros}

\begin{abstract}
Tensor algebra is essential for data-intensive workloads in various computational domains. Computational scientists face a trade-off between the specialization degree provided by dense tensor algebra and the algorithmic efficiency that leverages the structure provided by sparse tensors. This paper presents StructTensor, a framework that symbolically computes structure at compilation time. This is enabled by Structured Tensor Unified Representation (STUR), an intermediate language that can capture tensor computations as well as their sparsity and redundancy structures. Through a mathematical view of lossless tensor computations, we show that our symbolic structure computation and the related optimizations are sound. Finally, for different tensor computation workloads and structures, we experimentally show how capturing the symbolic structure can result in outperforming state-of-the-art frameworks for both dense and sparse tensor algebra.
\end{abstract}

\maketitle

\section{Introduction}

\begin{figure}
    \centering
    \includegraphics[width=0.85\linewidth]{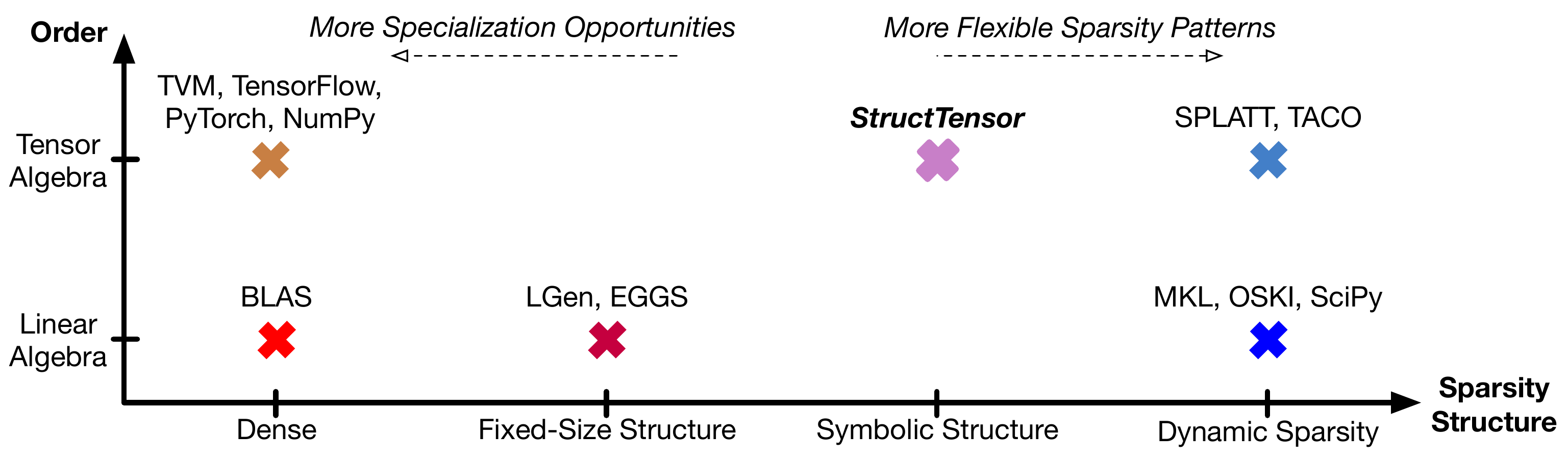}
    \caption{Comparison of different linear and tensor algebra frameworks.}
    \label{fig:tizer}
\end{figure}

Linear and tensor algebra operations are the key drivers of data-intensive computations in many domains, such as physics simulations~\cite{martin2008xperm,ran2020tensor}, computational chemistry~\cite{titov2013generating,hirata2006symbolic}, bioinformatics~\cite{cichocki2009nonnegative}, and deep learning~\cite{DBLP:journals/jossw/SmithG18,doi:10.1021/jp034596z}. 
Due to their importance, many specialization attempts have been made throughout the entire system stack from hardware to software layers. The tensor accelerators~\cite{DBLP:conf/micro/HegdeMPCJSEF19} and TPUs~\cite{DBLP:conf/isca/JouppiYPPABBBBB17} provide efficient tensor processing at the hardware level. On the software level, there have been advances in providing highly-tuned kernels~\cite{DBLP:journals/toms/DongarraCHD90}, as well as compilation frameworks that globally optimize tensor computations~\cite{DBLP:journals/taco/GareevGK18}.

There is a trade-off for using tensor algebra frameworks between the specialization degree and the flexibility of the structure such as sparsity (Figure~\ref{fig:tizer}). 
On the one side of the spectrum, extensive research has been done on dense tensors without leveraging any structure. 
Such tensors appear in deep neural networks and computational physics. As all the memory access patterns are known at compilation time, one can provide heavily-tuned implementations without any knowledge about the content of tensors. As a result, the high-performance engineer or compiler has enough reasoning power to make decisions on parallelization, vectorization, and tiling. 

However, many real-world applications involve tensors that exhibit symmetries and, more generally, specific structures which can be exploited to significantly reduce computational costs. 
Sparse tensor algebra for instance only captures the pattern of zero/non-zero elements in tensors during runtime, and has been at the center of recent interest~\cite{strout2018sparse,kjolstad:2017:taco}. 
However, postponing the memory access patterns to the runtime hinders the specialization power of the compiler~\cite{augustine2019generating}.
As a partial remedy, there have been efforts~\cite{DBLP:journals/cgf/TangSKPLP20,spampinato2016basic} to statically determine the structure of matrices during the compilation time. However, these are limited to fixed-size matrices.

To resolve the dilemma between using tensor algebra frameworks focusing on either dense or sparse tensors (Figure~\ref{fig:tizer}), this paper introduces \systemname{}. \systemname{} captures the structure of tensors \textit{symbolically} at compilation time. On the one hand, the underlying compiler can use this symbolic information to specialize the code at the level of dense computations. On the other hand, the compiler can leverage this symbolic information in order to eliminate unnecessary and redundant computation. 

In \systemname{}, all tensor computations and structure information are translated to a single intermediate language called Structured Tensor Unified Representation (\unifiedir{}). \unifiedir{} propagates the structure knowledge throughout the computation at compile time. This is followed by efficient C++ code generation.

Specifically, we make the following contributions:
\begin{itemize}
    \item We present \systemname{}, the first framework that supports structured computation for tensor algebra (Section~\ref{sec:overview}). For a tensor $T$, the structure handled by \systemname{} comes as a pair $(T_U,T_R)$, where
    \begin{itemize}
        \item $T_U$ tracks the symbolic sparsity structure of the tensor,
        \item $T_R$ tracks the redundancy structure, which captures symmetries and repetition patterns.
    \end{itemize}
    \item We propose \unifiedir{}, a unified intermediate representation (IR) that can express tensor computations. It allows capturing non-optimized tensors as well as their symbolic structures $T_U, T_R$ in a single IR (Section~\ref{sec:system}). 
    \item We show how \systemname{} uses the structure information in its compilation process (Section~\ref{sec:compile}). It leverages \unifiedir{} and rewrites tensors in 3 steps (also see Figure~\ref{fig:architecture}): 
    \begin{enumerate}
        \item \systemname{} propagates the structure information on the syntax tree of tensor computation (Section~\ref{sub:structure-inference})
        \item Several optimizations are applied to the unified representation for tensor computations and their structure leading to computation over a compressed tensor (Section~\ref{sub:optimizations})
        \item \systemname{} generates C++ code for the tensor computation over the compressed form, as well as reconstructing the uncompressed tensor (Section~\ref{sub:code-generation})
    \end{enumerate}
    \item We give a mathematical view on the problem of lossless tensor computations that we study in this paper, and show the soundness of our rewrites and structure inference in Section~\ref{sec:soundness}.
    \item We experimentally evaluate \systemname{} for different tensor computation workloads and structures (Section~\ref{sec:exp}). We show that \systemname{} can leverage the structure of tensors to generate computation over compressed tensors in order to outperform state-of-the-art dense and sparse tensor computation frameworks.
\end{itemize}

\section{Background}
\label{bg}

\smartpara{Structured Linear Algebra}
There are several well-known structures for matrices, such as diagonal, symmetric, and lower/upper triangular, that considering them while doing calculations over matrices can reduce computational time. Figure~\ref{fig:structla} represents a set of linear algebra operations and well-known matrix structures. 

\examplepara{Diagonal Matrices}
Imagine the case of the Kronecker product between two diagonal matrices $A_{n \times n}$ and $B_{m \times m}$. Typically, the Kronecker product of these matrices is defined as: 
\begin{equation}
(A \otimes B)_{mr+v, ms+w} = A_{r,s} \cdot B_{v,w}
\end{equation}
where the result dimension is $nm \times nm$. Therefore the computational cost would be $O(n^2 m^2)$. However, if the computation is structure-aware, one can leverage the fact that all non-zero elements are on the diagonal of matrices. Therefore, it is sufficient to perform the multiplication over only those elements, which results in the following computation:
\begin{equation}
(A \otimes B)_{mr+v, mr+v} = A_{r,r} \cdot B_{v,v}
\end{equation}
where the result is diagonal as well. As a result, the computational complexity for this Kronecker product reduces to $O(nm)$. Furthermore, since the structure of the result is known (diagonal), this information can be used in further computations efficiently. Figure~\ref{fig:structla} shows a subset of inference rules to determine the output structure based on input structures.

\input{figures/grammar-la}

\smartpara{Sparsity and Redundancy Structures}
Structures that distinguish zero and non-zero values are referred to as \textit{sparsity structures}, such as diagonal and lower/upper triangular. Other structures like symmetric can capture redundancy patterns and are thus referred to as \textit{redundancy structures}. Because of the redundancy pattern, it is possible to reconstruct the whole matrix by storing half of the data even though they are non-zero. Knowing and propagating structures in many cases helps reduce computational costs~\cite{spampinato2016basic}.

\smartpara{Intricate Structures}
The aforementioned structures cannot cover more complicated patterns. Imagine the case of creating the covariance matrix for the polynomial regression degree two model.\footnote{We consider factorization-related optimizations, techniques that aim to improve the performance by leveraging low ranks of tensors, orthogonal to structured tensors; our technique appears after such transformations, as can be seen in Section~\ref{sec:exp:e2e}.}
Figure~\ref{fig:PR2Covar} shows the covariance matrix created for a polynomial degree two model on a data element with 4 features. Each number and color is associated with distinct elements, and elements with the same number have the same value. As it is represented in Figure~\ref{fig:PR2Covar}, there are only 65 distinct elements even though the covariance matrix dimension is $20 \times 20$. The redundancy pattern in this matrix is sophisticated and cannot be captured using the existing structures. Utilizing such structure information enhances the performance of machine learning tasks that require covariance matrix creation such as training polynomial regression models.

\smartpara{Tensor Structures}
The complicated structure of the mentioned covariance matrix can be captured in the form of a higher-order tensor structure. 
This covariance matrix contains all degree-2, degree-3, and degree-4 interactions between features in itself. 
The reason that redundancy happens is that these interactions are computed multiple times. 
For example, degree-2 interactions inside the matrix form a symmetric structure. 
Degree-3 and degree-4 interactions have some complex forms of redundancy as well as being symmetric. 
These interactions cover all the elements inside the covariance matrix. By representing degree $i$ interactions as a tensor of order-$i$, the computation can be done more efficiently. This way, the only unique elements in degree two, three, and four tensors $T2(i, j), T3(i, j, k), T4(i, j, k, l)$ reside in $\{ T2(i, j) \mid 1 \leq i \leq j \leq n \}, \{ T3(i, j, k) \mid 1 \leq i \leq j \leq k \leq n \}, \{ T4(i, j, k, l) \mid 1 \leq i \leq j \leq k \leq l \leq n \}$ respectively. This structure in the new data format forms a generalized symmetric pattern that is easy to capture and leads to efficient computation. In the following sections, it is explained how \systemname{} handles sophisticated structures and generates efficient C++ code for the computations.
\section{Overview}
\label{sec:overview}

\begin{figure}
    \centering
    \includegraphics[width=\linewidth]{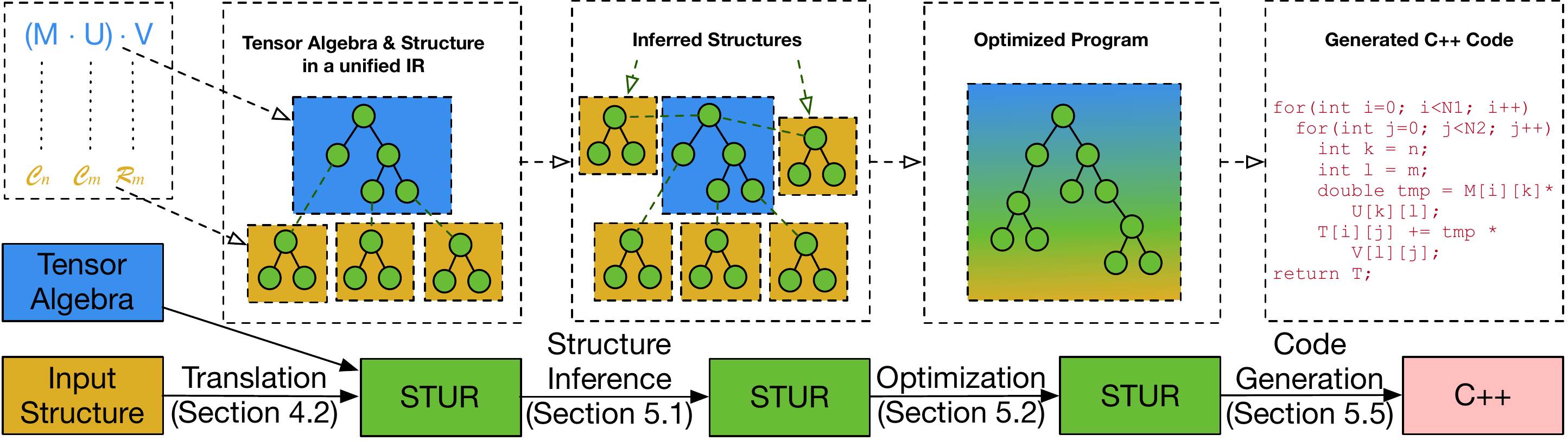}
    \caption{\systemname{} architecture overview.}
    \label{fig:architecture}
\end{figure}

In this section, the overall architecture of \systemname{} is described (cf. Figure~\ref{fig:architecture}).

\smartpara{Input}
\systemname{} gets linear algebra and tensor algebra expressions and the structure of their parameters as input. These operations and structures subsume the ones that are mentioned in Figure~\ref{fig:structla}. All tensor expressions, as well as their structures, are represented in the \systemname{} unified representation, called \unifiedir{}. Sparsity and redundancy structures are captured in a unique set and redundancy map, respectively. The unique set contains unique and non-zero elements, and the redundancy map has the mapping from redundant and non-zero elements outside the unique set to their corresponding value in the unique set. These two subsume \texttt{SInfo} and \texttt{AInfo} used in LGen~\cite{spampinato2016basic} to model structured matrices.

\smartpara{Structure Inference}
Afterward, we infer the structure of the intermediate and output tensor expressions by propagating the structure through \unifiedir{}. A predefined set of inference rules for operations and structures is provided in \unifiedir{}. The output structure is inferred in terms of the unique set and redundancy map by using inference rules. The inference rules set is extensible to cover arbitrary operations and structures as well.

\smartpara{Optimizations}
Various optimizations are applied to \unifiedir{} expressions. Throughout the structure inference process, several intermediate unique sets and redundancy maps are created. The tensor inlining optimization removes intermediate sets and maps. Moreover, input structures and previous optimizations by \unifiedir{} can produce repetitive conditions over iterators. Through logical simplifications, only distinct conditions are kept for the code generation step. 

\smartpara{Code Generation}
Finally, the optimized \unifiedir{} is fed to the C++ code generator. The code generator assumes an order for iterators, calculates loop nest boundaries, and generates an efficient structure-aware C++ code. The generated code performs the computations over compressed format with lower computational cost thanks to the unique set. By utilizing the redundancy map, the final output tensor can be reconstructed for the user. 

\examplepara{Polynomial Regression Degree-2}
We consider the creation of a covariance matrix for polynomial regression degree-2.
The covariance matrix for an element with $n$ features is defined as: 

\begin{center}
$\Sigma(\mathbf{x}) = \mathbf{x} \cdot \mathbf{x}^T = \mathbf{x} \otimes \mathbf{x}$
\end{center}

\noindent
This represents the outer product of $\mathbf{x}$ with itself, and thus can be represented using the vector outer product operation $\otimes$.
The vector $\mathbf{x}$ is a vector resulting from the concatenation of the feature vector $\mathbf{f}$ with the vector resulting from the interaction of the features. 
In terms of linear algebra, this is expressed as:

\begin{center}
$\mathbf{x} = \mathbf{f} \oplus \vectorize{\mathbf{f} \otimes \mathbf{f}}$
\end{center}

\noindent Here, $\oplus$ is the vector concatenation, $\otimes$ is the vector outer product, and $\vectorize{M}$ flattens the matrix $M$ of size $n\times m$ into a vector of size $n\cdot m$. 
By inlining the definition of $\textbf{x}$ and by using the identity for the distribution of vector concatenation and vector outer product, we have:

\begin{center}

\begin{tabular}{r c r c l}
$\Sigma(\mathbf{x})$ & $=$ & $(\mathbf{f} \oplus \vectorize{\mathbf{f} \otimes \mathbf{f}})$ & $\otimes$ & $(\mathbf{f} \oplus \vectorize{\mathbf{f} \otimes \mathbf{f}})$ \\
& $=$ & $\big( (\mathbf{f} \otimes \mathbf{f}) \hcat (\mathbf{f} \otimes \vectorize{\mathbf{f} \otimes \mathbf{f}})\big)$
& $\vcat$ & $\big( (\vectorize{\mathbf{f} \otimes \mathbf{f}} \otimes \mathbf{f}) \hcat (\vectorize{\mathbf{f} \otimes \mathbf{f}} \otimes \vectorize{\mathbf{f} \otimes \mathbf{f}})\big)$
\end{tabular}
\end{center}

\begin{figure}
    \centering
    \begin{tabular}{cc}
        \includegraphics[width=0.45\textwidth]{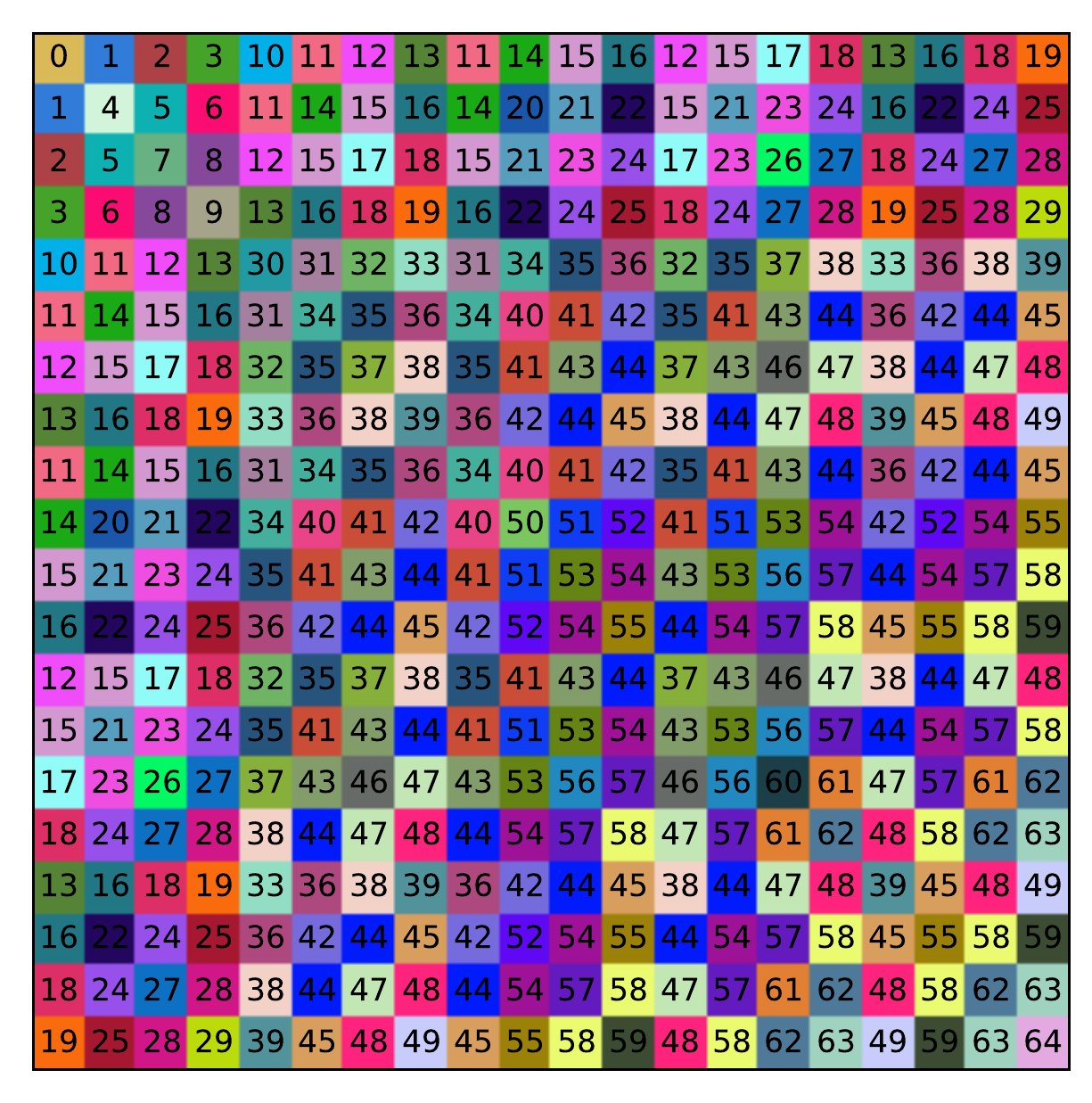} & 
        \includegraphics[width=0.45\textwidth]{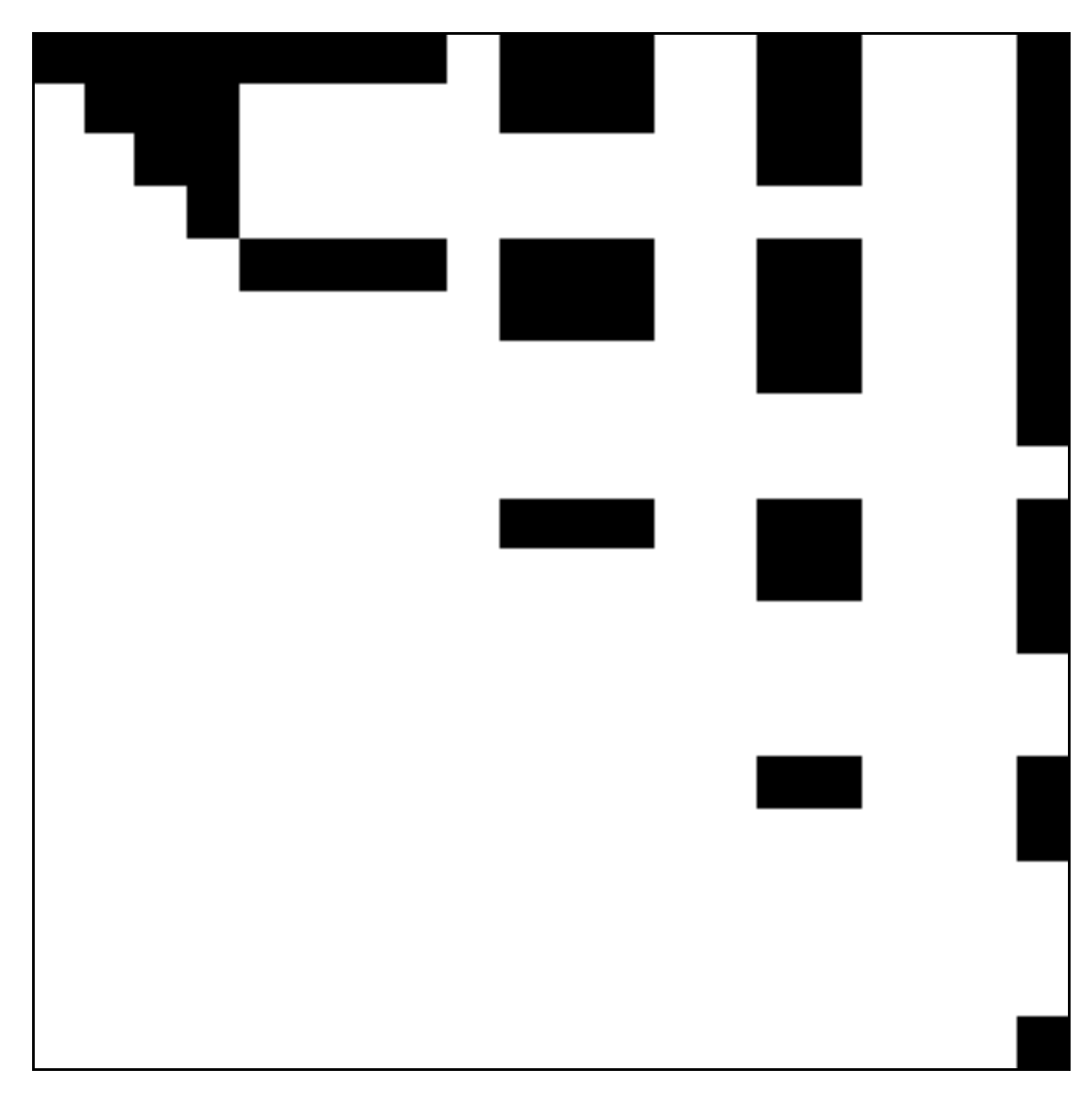}
    \end{tabular}
    \caption{The covariance matrix of Polynomial Regression degree-2. Its redundancy pattern is shown on the left and its unique elements in compressed format are on the right.}
    \label{fig:PR2Covar}
\end{figure}

\noindent The operations $\hcat$ and $\vcat$ correspond to matrix horizontal and vertical concatenation, respectively. \systemname{} improves the performance for this computation in two levels of granularity. In a coarse-level of granularity, it detects that $(\mathbf{f} \otimes \vectorize{\mathbf{f} \otimes \mathbf{f}})$ and $(\vectorize{\mathbf{f} \otimes \mathbf{f}}\otimes \mathbf{f})$ are computing the same elements but in a different layout. Thus, it only performs one of the computations and it would be sufficient to compute the following vector outer product terms:

\begin{center}

\begin{tabular}{r c l}
$M1$ & $=$ & $\mathbf{f} \otimes \mathbf{f}$ \\
$M2$ & $=$ & $\mathbf{f} \otimes \vectorize{\mathbf{f} \otimes \mathbf{f}}$ \\
$M3$ & $=$ & $\vectorize{\mathbf{f} \otimes \mathbf{f}} \otimes \vectorize{\mathbf{f} \otimes \mathbf{f}}$ \\
\end{tabular}
\end{center}

\noindent In a finer-level of granularity, for each of the terms $M1, M2, M3$, \systemname{} detects a generalized symmetric structure.
For $M1$ it detects a standard symmetric structure where it is sufficient to only keep the upper half of the matrix.
For $M2$ and $M3$, there are $\sim 5\times$ and $\sim 23\times$ redundant elements, the patterns of which can be seen in Figure~\ref{fig:PR2Covar}. After inferring such structures, \systemname{} generates the following C++ code for each of these terms:

\noindent
\begin{tabular}{l|l|l}
        \begin{lstlisting}[language=C++, basicstyle=\scriptsize]
// Computation for M1
for(int i=0; i<n; ++i){
 for(int j=i; j<n; ++j){
  M1[i][j]=f[i]*f[j];
}}
        \end{lstlisting} & 
        \begin{lstlisting}[language=C++, basicstyle=\scriptsize]
// Computation for M2
for(int i=0; i<n; ++i){
 for(int j=i; j<n; ++j){
  for(int k=j; k<n; ++k){
   int col=j*n+k;
   M2[i][col]=f[i]*f[j]*f[k];
}}}
        \end{lstlisting} & 
                \begin{lstlisting}[language=C++, basicstyle=\scriptsize]
// Computation for M3
for(int i=0; i<n; ++i){
 for(int j=i; j<n; ++j){
  for(int k=j; k<n; ++k){
   for(int l=k; l<n; ++l){
    int r=i*n+j;
    int c=k*n+l;
    M3[r][c]=f[i]*f[j]*f[k]*f[l];
}}}}
        \end{lstlisting}
\end{tabular}

\noindent Note that the nested iterations only cover a subset of the range.

In the next sections, we will thoroughly elaborate on how \systemname{} works.

\section{\unifiedir{}: Structured Tensor Unified Language} 
\label{sec:system}

In this section, the syntax of the unified intermediate representation, \unifiedir{}, is represented. We elaborately explain the grammar for the unique set, redundancy map, and compressed tensor computation in this section. Moreover, details of how structured linear algebra is represented in \unifiedir{} can be found in this section.

\input{figures/grammar-core}

\subsection{The syntax of \unifiedir{}}
\smartpara{Grammar}
Figure~\ref{fig:grammar} shows \unifiedir{} grammar which covers the grammar for tensor, unique set, and redundancy map computations. Each program (\metavar{P}) is made of several rules (\metavar{R}). Each rule is in the form of an assignment from a body (\metavar{B}) to an access to a collection (\metavar{A}). 
The collection could be a tensor (\metavar{T}), compressed tensor (\metavar{T}$_C$), unique set (\metavar{T}$_U$), or redundancy map (\metavar{T}$_R$), whereas the access index can be multiple index variables (\metavar{\overline{X}}), an index variable, or a constant value.
The assignment body is represented as a sum of factor products (\metavar{F}). Each factor restricts the domain of values that an index variable can have. This is achieved through a collection access or a comparison term. 

\examplepara{Simple Tensor Operation} Consider the following \unifiedir{} program:

\begin{center}
\begin{tabular}{r c l}
     $T_3(x, y)$ & $:=$ & $T_1(x, y) * T_2(x, y)$\\
     $T_4(x)$ & $:=$ & $T_3(x, y) * (x = y)$
\end{tabular}
\end{center}

\noindent This program consists of two rules, with two input tensors $T_1$ and $T_2$. 
The first rule constructs the tensor $T_3$ of order-2, i.e., a matrix, which is computed by performing an element-wise multiplication of the two input matrices. The second rule constructs an order-1 tensor, i.e., a vector, that contains the elements of the diagonal of the matrix $T_3$. This is achieved by (1) restricting the range of $x$ and $y$ by the comparison term $x=y$, and (2) existentially quantifying over $y$ by not including it in the head of the rule.

\smartpara{Sum-of-Product Semantics}
The addition and multiplication in \unifiedir{} are defined based on the underlying collection; for unique sets and redundancy maps, the addition and multiplication are defined as set union and intersection, whereas for tensors and compressed tensors they are defined as real number addition and multiplication. Each rule can existentially quantify the free variables of the body by not including them in the rule head. This is known as \textit{marginalization} in the AI community. 
All the free variables in the head should already be defined in the body (cf. Figure~\ref{fig:freevars})

\smartpara{Syntactic Sugar} To simplify the presentation, we consider the following syntactic sugar, where $\argvec{x}$ corresponds to a list of arguments $x_1, \ldots, x_k$: 

\begin{center}
\begin{tabular}{r c l}
  $(a \leq x < c)$   & $\equiv$ &  $(a \leq x) * (x < c)$ \\
  $(x = y = z)$   & $\equiv$ & $(x = y) * (y = z)$  \\
  $\argvec{x} \theta \argvec{b}$ & $\equiv$ & $(x_1 \theta b_1) * \ldots * (x_k \theta b_k) $
\end{tabular}
\end{center}

\input{figures/freevars}

\smartpara{Unique sets} The sparsity structure of all the distinct elements of a tensor is encoded in a unique set. The grammar of \unifiedir{} already captures the definition of unique sets (cf. Figure~\ref{fig:grammar}); a unique set is provided as a sum of products of comparison terms or access to other unique sets (and redundancy maps). 
Unique sets enhance performance by restricting index boundaries. 

\examplepara{Chess Pattern Unique Set} The following \unifiedir{} specifies the unique set of a matrix $T$ of size $m\times n$ with a chess-board pattern for the sparsity:

\begin{center}
\begin{tabular}{r c l}
     $T_U(i, j)$ & $:=$ & $(0 \leq i' < (m / 2)) * (0 \leq j' < (n / 2)) * (i = i' * 2) * (j = j' * 2 + 1)$ $ + $ \\
     && $(0 \leq i' < (m / 2)) * (0 \leq j' < (n / 2)) * (i = i' * 2 + 1) * (j = j' * 2)$
\end{tabular}
\end{center}

\noindent The first factor product specifies the elements of the even rows (that have an odd column index), and the second one specifies the ones for odd rows (with an even column index).

\smartpara{Redundancy Maps} The remaining non-zero elements which do not appear in the unique set domain are covered by the redundancy map. A redundancy map keeps the association between these elements' indices and their corresponding indices in the unique set. Similar to the unique set, it is represented as a sum of products of comparison terms or access to other unique sets and redundancy maps. For a tensor of order-$k$, the redundancy map has $2k$ index variables. The first $k$ index variables correspond to the indices of the redundant element, whereas the second $k$ one corresponds to the indices from the unique set. 
The redundancy map enables the capability of reconstructing the full matrix from the compressed version that only contains unique elements. Restricting the computations to the elements of the unique set and reconstructing the uncompressed final result once, when all the calculation is over, can significantly improve the performance.

\examplepara{Identical Row Matrix Redundancy Map} Consider a matrix $T$ of size $m\times n$, where all rows are the same. The unique set and redundancy maps for this matrix are defined as follows:

\begin{center}
\begin{tabular}{r c l}
     $T_U(i, j)$ & $:=$ & $(i = 0) * (0 \leq j < n)$ \\
     $T_R(i, j, i', j')$ & $:=$ & $(0 < i < m) * (0 \leq j < n) * (i' = 0) * (j' = j)$ \\
\end{tabular}
\end{center}

\noindent The first rule specifies that the unique elements are only in the first row ($i = 0$). The first term of the second rule $(0 < i < m)$ restricts the range of redundant elements to the ones from all the rows except the first one. The last two terms specify the index of the corresponding element from the unique set by specifying the first row ($i'=0$) and the same column as the redundant element ($j'=j$).

\smartpara{Compressed Tensor} \systemname{} leverages the structure for better performance by representing the tensor in a lossless compressed format. This is achieved by combining the original tensor with the unique set in order to extract only the unique elements. The compressed tensor $T_C$ for the original tensor $T$ is defined as: 

\begin{center}
    $T_C(\argvec{x}) := T(\argvec{x}) * T_U(\argvec{x})$
\end{center}

\noindent Na\"ively executing the computation using this formula can even make the performance worse. After performing simplifications, the code generator produces code that iterates over the domain provided by the unique set for operations and excludes all the other elements of that tensor. This way, no extra computational cost is imposed while computing the result. When the computation is done, the uncompressed tensor $T$ is retrievable by using $T_R$.

\examplepara{Upper Triangular Matrix Compressed Tensor} A $n \times n$ upper triangular matrix compressed tensor is calculated as follows:

\begin{center}
\begin{tabular}{r c l}
$T_U(i, j)$ & $:=$ & $(0 \leq i \leq j < n)$\\
$T_C(i, j)$ & $:=$ & $T(i, j) * (0 \leq i \leq j < n)$
\end{tabular}
\end{center}

\noindent When $T(i, j)$ appears in a computation, the optimizer converts it to $T_C(i, j)$ and only uses elements with $0 \leq i \leq j < n$. Therefore, only half of the elements are used in the computation, which leads to a $\sim2\times$ speed up.

\input{figures/la-in-stur}

\input{figures/structure-in-stur}

\subsection{Translating Structured Linear Algebra to \unifiedir{}}

\smartpara{Operations}
The representation of linear algebra operations is shown in Figure~\ref{fig:operationIR}. Imagine the matrices $M_1$ and $M_2$ with dimensions $m_1 \times n_1$ and $m_2 \times n_2$ are represented in \unifiedir{} as \translate{$e_1$} and \translate{$e_2$}, respectively. The element in $i$-th row and $j$-th column of $M_1 \sbullet M_2$ where $\sbullet$ is an operator is \translate{$e_1 \sbullet e_2$}$(i, j)$. Each operation is translated to a sum of product format in \unifiedir{}. For instance, if $\sbullet = \oplus$, then the direct sum of $M_1$ and $M_2$ is translated by the top-right rule in Figure~\ref{fig:operationIR}. This rule specifies that element $(i, j)$ of the output is taken from \translate{$e1$} when $i < m$ and $j < n$ and from \translate{$e2$} when $i \geq m$ and $j \geq n$.

\smartpara{Structures}
Figure~\ref{fig:structureIR} shows the representation of well-known matrix structures in \unifiedir{}. Structures are translated to a combination of a unique set and a redundancy map. For example, if matrix $M$ has a symmetric structure, the upper triangular section of that is counted as the unique set. Therefore, the lower triangular region is considered as the redundant section; the redundancy map keeps the transformations from the lower triangular region to the corresponding upper region. Hence, if the representation of $M$ in \unifiedir{} is $T$, the unique set and redundancy map are provided by the last rule of Figure~\ref{fig:structureIR}. When $j<i$, according to $T_R(i, j, i', j')$, the values of $T(i, j)$ will be retrieved from values in $T(j, i)$.

\section{Compilation}
\label{sec:compile}
In this section, we show how \systemname{} symbolically computes and propagates the structure. 
Optimizing the tensor computation using the inferred structure leads to structure-aware code generation. 

\subsection{Structure Inference}
\label{sub:structure-inference}

\input{figures/infer-unqiueset}

\smartpara{Inference for unique sets}
Figure~\ref{fig:uniqueset} shows the inference rules for the output structure. These rules follow the assumption that input redundancy maps are empty and only capture sparsity patterns. Since all operations are translated to multiplication, addition, or projection in \unifiedir{}, providing inference rules only for these operations is sufficient. 
Moreover, having more than two operands is handled by breaking them into subexpressions with two operands and storing the results in intermediate sets. 

In the case of multiplication, the result is non-zero only if both inputs are non-zero values. Having a zero operand in multiplication makes the final result zero. Therefore, the unique set of output is calculated from the intersection of non-zero elements in both operands. Addition on the other hand is non-zero even if one of the operands is non-zero. Hence, any element in the unique set of operands can lead to a non-zero element in output. Consequently, the unique set of output is the union over the unique set of its operands. 
Since projection is defined by summing up all values, it follows the same rule as addition.
Any non-zero element in the unique set of the input can create a non-zero element in the output. So the output unique set for projection is computed by unioning over all unique set values in that dimension.

\examplepara{Unique Set Computation} Consider the following tensor computation:

\noindent
\begin{center}
\begin{tabular}{r c l}
 $A(i, j, k)$ & $:=$ & $B(i, j, l) * C(i, k) + D(j, k) * E(i, j)$
\end{tabular}
\end{center}

\noindent
where input unique sets are 

\noindent
\begin{tabular}{r c l r c l}
$B_U(i, j, l)$ & $:=$ & $i \leq j,$ & $C_U(l, k)$ & $:=$ & $l = k$ \\
$D_U(j, k)$ & $:=$ & $ \emptyset,$ & $E_U(i, k)$ & $:=$ & $i \leq k$ 
\end{tabular}

\noindent
Output unique set computation is expressed as follows:

\noindent

\begin{tabular}{r c l r c l}
$S_U(i, j, k, l)$ & $:=$ & $B_U(i, j, l) * C_U(l, k)$, &
$P_U(i, j, k)$ & $:=$ & $S_U(i, j, k, l)$ \\
$Q_U(i, j, k)$ & $:=$ & $D_U(j, k) * E_U(i, k)$,  &
$A_U(i, j, k)$ & $:=$ & $P_U(i, j, k) * Q_U(i, j, k)$ \\
\end{tabular}

\input{figures/infer-redundancy}

\smartpara{Inference for redundancy maps}
The redundancy map of the output is required in order to have access to every element in the output tensor and reconstruct it. Figure~\ref{fig:redundancymap} shows the rules to infer output redundancy map and unique sets. Rules for the tensor outer product, addition, direct sum, and repetition are provided. For the cases where there is no rule in Figure~\ref{fig:redundancymap}, first, the redundancy maps of inputs are set to empty by combining all non-zero elements from the unique set and redundancy map. Then, rules from Figure~\ref{fig:uniqueset} are applied to the computation to calculate their unique set.

\examplepara{Structure Inference} Assume the following definition:

\begin{center}
$T(x_1, x_2, x_3, x_4, x_5) := M(x_1, x_2, x_3) * V(x_4, x_5)$
\end{center}

\noindent
Furthermore, consider the dimensions as $(d_1, d_2, d_3)=dims(M)$ and $(d_4, d_5)=dims(V)$, and the unique set and redundancy map of inputs are:

\noindent
\begin{tabular}{r c l}
$M_U(x_1, x_2, x_3)$ & $:=$ & $(x_1 > x_2) * (0 \leq x_1 < d_1) * (0 \leq x_2 < d_2) * (0 \leq x_3 < d_3)$ \\
$V_U(x_4, x_5)$ & $:=$ & $(x_4 \leq x_5) * (0 \leq x_4 < d_4) * (0 \leq x_5 < d_5)$ \\
$M_R(x_1, x_2, x_3, x'_1, x'_2, x'_3)$ & $:=$ & $\emptyset$ \\
$V_R(x_4, x_5, x'_4, x'_5)$ & $:=$ & $(x_4 > x_5) * (0 \leq x_4 < d_4) * (0 \leq x_5 < d_5) * (x'_4 = x_5) * (x'_5 = x_4)$
\end{tabular}

\noindent
Then output unique set and redundancy map are:

\noindent
\begin{tabular}{r c l}
$T_U(x_1, x_2, x_3, x_4, x_5)$ & $:=$ & $M_U(x_1, x_2, x_3) * V_U(x_4, x_5)$ \\
$T_R(x_1, x_2, x_3, x_4, x_5,$ & $:=$ & $M_R(x_1, x_2, x_3, x'_1, x'_2, x'_3) * V_R(x_4, x_5, x'_4, x'_5)$ $+$\\
$x'_1, x'_2, x'_3, x'_4, x'_5)$ & & $M_U(x_1, x_2, x_3) * V_R(x_4, x_5, x'_4, x'_5)$ $+$\\
& & $M_R(x_1, x_2, x_3, x'_1, x'_2, x'_3) * V_U(x_4, x_5)$\\
\end{tabular}

\smartpara{Extensions}
The provided set of rules in Figures~\ref{fig:uniqueset} and ~\ref{fig:redundancymap} are for now relatively minimal and are meant to be extensible. 

For example, the rule for self-multiplication can be easily extended to consider higher-degree self-multiplications. The idea is that our rewriting system will always choose the most specialized (and therefore optimizing) rule first.

\subsection{Optimizations}
\label{sub:optimizations}

\smartpara{Rule Inlining}
The intermediate tensors, unique sets, and redundancy maps are materialized. 
Inlining these definitions can improve performance in various ways. 
First, one can avoid the materialization overhead. 
Second, the inlining is an enabling transformation that can open opportunities for further optimizations.
During the inlining, the tensor index variable needs to be alpha renamed in order to avoid capturing free variables.

\smartpara{Logical Simplifications} After inlining, the factors inside unique sets and redundancy maps might be repetitive or result in $\emptyset$ or other simple rules. Logical simplification removes the repetitive conditions. Furthermore, there could be contradicting conditions that will result in $\emptyset$ (e.g., $(a \leq b) * (a > b)$ lead to $\emptyset$).
Multiplying by $\emptyset$ results in $\emptyset$ (i.e., $\emptyset * e = \emptyset$), and it is the neutral element for addition (i.e., $\emptyset + e = e$).

\examplepara{Structure Optimization} Consider the previous example. After inlining, the unique set and redundancy map are transformed as follows:

\noindent
\begin{tabular}{r c l}
$T_U(x_1, x_2, x_3, x_4, x_5)$ & $:=$ & $(x_1 > x_2) * (0 \leq x_1 < d_1) * (0 \leq x_2 < d_2) * (0 \leq x_3 < d_3)$ $*$\\
& & $~~(x_4 \leq x_5) * (0 \leq x_4 < d_4) * (0 \leq x_5 < d_5)$\\
$T_R(x_1, x_2, x_3, x_4, x_5,$ & $:=$ & $\emptyset * (x_4 > x_5) * (0 \leq x_4 < d_4) * (0 \leq x_5 < d_5) * (x'_4 = x_5) * (x'_5 = x_4)$ $+$\\
$x'_1, x'_2, x'_3, x'_4, x'_5)$ & & $(x_1 > x_2) * (0 \leq x_1 < d_1) * (0 \leq x_2 < d_2) * (0 \leq x_3 < d_3)$ $*$\\
& & $~~~(x_4 > x_5) * (0 \leq x_4 < d_4) * (0 \leq x_5 < d_5) * (x'_4 = x_5) * (x'_5 = x_4)$ $+$\\
& & $\emptyset * (x_4 \leq x_5) * (0 \leq x_4 < d_4) * (0 \leq x_5 < d_5)$\\
\end{tabular}

\noindent By further simplifying the redundancy map by propagating the rules for $\emptyset$, we have:

\noindent
\begin{tabular}{r c l}
$T_R(x_1, x_2, x_3, x_4, x_5,$ & $:=$ & $(x_1 > x_2) * (0 \leq x_1 < d_1) * (0 \leq x_2 < d_2) * (0 \leq x_3 < d_3)$ $*$\\
$x'_1, x'_2, x'_3, x'_4, x'_5)$ & & $~~~(x_4 > x_5) * (0 \leq x_4 < d_4) * (0 \leq x_5 < d_5) * (x'_4 = x_5) * (x'_5 = x_4)$\\
\end{tabular}

\subsection{Use case 1: structured linear algebra}

In this section, we show that \systemname{} can recover the output structure of linear algebra operations by using the inference and optimization rules presented earlier.
Next, we provide two examples.

\examplepara{Upper-triangular Hadamard product by Symmetric (UHS)} Assume two square matrices $M$ and $N$ with dimension $n\times n$ and with upper triangular and symmetric structures, respectively. The element-wise multiplication is represented in \unifiedir{} as follows:

\begin{center}
    $A(i, j) := M(i, j) *  N(i, j)$
\end{center}
The unique sets and redundancy maps of the input matrices are represented as follows:

\noindent
\begin{tabular}{r  c l}
$M_U(i, j)$ & $:=$ & $(0 \leq i \leq j < n)$ \\
$N_U(i, j)$ & $:=$ & $(0 \leq i \leq j < n)$ \\
$M_R(i, j, i', j')$ & $:=$ & $\emptyset$ \\
$N_R(i, j, i', j')$ & $:=$ & $(0 \leq j < i < n) * (i' = j) * (j' = i)$
\end{tabular}

The output unique set will be:

\noindent
\begin{tabular}{r c l}
$A_U(i, j)$ & $:=$ & $M_U(i, j) * N_U(i, j) + M_U(i, j) * N_R(i, j, i', j') + M_R(i, j, i', j') * N_U(i, j)$ \\
(inlining) & $:=$ & $\big((0 \leq i \leq j < n) * (0 \leq i \leq j < n)\big)$\\
& & $ + \big((0 \leq i \leq j < n) * (0 \leq j < i < n) * (i' = j) * (j' = i)\big)$\\
& & $ + \big(\emptyset * (0 \leq i \leq j < n)\big)$\\
(simplification) & $:=$ & $(0 \leq i \leq j < n) + \emptyset + \emptyset$\\
& $:=$ & $(0 \leq i \leq j < n)$ \\
\end{tabular}

\examplepara{Row matrix multiplied by Diagonal (RMD)} Consider an $m \times n$ row matrix $M$, which only has elements in its $r$-th row, that is being multiplied by an $n \times n$ diagonal matrix $N$. The computation is expressed as:

\begin{center}
$A(i, j) := M(i, k) * N(k, j)$
\end{center}

\noindent
The unique set and redundancy maps are as follows:

\noindent
\begin{tabular}{r c l l}
$M_U(i, k)$ & $:=$ & $(i = r) * (0 \leq j < n),$ & $M_R(i, k, i', k') := \emptyset$\\
$N_U(k, j)$ & $:=$ & $(k = j) * (0 \leq k < n),$ & $N_R(k, j, k', j') := \emptyset$
\end{tabular}

Since there is no redundancy map, Figure~\ref{fig:uniqueset} rules are enough to infer the output unique set. Output redundancy map is naturally $\emptyset$. Therefore, the output unique set is:

\noindent
\begin{tabular}{r c l}
$A_U(i, j)$ & $:=$ & $M_U(i, k) * N_U(k, j)$\\
(inlining) & $:=$ & $(i = r) * (0 \leq j < n) * (k = j) * (0 \leq k < n)$\\
(simplification) & $:=$ & $(i = r) * (0 \leq j < n) * (0 \leq j < n)$\\
(simplification) & $:=$ & $(i = r) * (0 \leq j < n)$
\end{tabular}

\noindent
As it is shown in Figure~\ref{fig:structureIR}, this unique set corresponds to a row matrix.

\subsection{Use case 2: structured tensor algebra} 
As explained, \systemname{} uses inference rules in \unifiedir{} and produces an optimized code for tensor algebra. Similar to the case of linear algebra, the programmer provides the tensor algebra program as input alongside the structure for input tensors. 
As there are massively many possibilities for the structure of higher-order tensors, we expect the programmer to provide them in \unifiedir{}.

\examplepara{Diagonal Tensor Times Vector (DTTV)}
In this example, we show how the output structure for TTV (Tensor Times Vector) is inferred in \unifiedir{}. TTV is defined as the following \unifiedir{} expression:

\begin{center}
$A(i, j) := M(i, j, k) * N(k)$
\end{center}

\noindent Consider the unique set and redundancy map of the inputs to be as follows:

\noindent
\begin{tabular}{r c l}
$M_U(i, j, k)$ & $:=$ & $(i = j) * (j = k) * (0 \leq i < m) * (0 \leq j < m) * (0 \leq k < m)$\\
$M_R(i, j, k, i', j', k')$ & $:=$ & $\emptyset,$ $\quad$ $V_U(k) := (0 \leq k < m),$ $\quad$ $V_R(k, k') := \emptyset$
\end{tabular}\\
where the dimensions are $(m, m, m) = dims(M)$ and $(m) = dims(V)$. Computation is divided into two steps, multiplication followed by reduction.

\noindent
\begin{tabular}{r c l}
$B(i, j, k)$ & $:=$ & $M(i, j, k) * N(k)$\\
$A(i, j)$ & $:=$ & $B(i, j, k)$\\
\end{tabular}\\
Therefore, the output unique set should be calculated in two steps as well.

\noindent
\begin{tabular}{r c l}
$B_U(i, j, k)$ & $:=$ & $M_U(i, j, k) * N_U(k)$\\
(inlining) & $:=$ & $(i = j) * (j = k) * (0 \leq i < m) * (0 \leq j < m) * (0 \leq k < m) * (0 \leq k < m)$\\
(simplification) & $:=$ & $(i = j) * (j = k) * (0 \leq i < m)$
\end{tabular}

\noindent
\begin{tabular}{r c l}
$A_U(i, j)$ & $:=$ & $B_U(i, j, k)$\\
(inlining) & $:=$ & $(i = j) * (j = k) * (0 \leq i < m)$\\
(simplification) & $:=$ & $(i = j) * (0 \leq i < m)$
\end{tabular}

\subsection{Code Generation}
\label{sub:code-generation}
As the final step, \systemname{} generates C++ code for the optimized \unifiedir{} expression.
To generate loop nests, first, the index variables are ordered following the same syntactic order as the input tensor expression\footnote{The optimal variable ordering is an NP-complete problem that is beyond the scope of this work.}. Afterward, \systemname{} computes the range of each index variable based on the output unique set. The range for each index variable is computed as the maximum of all its lower bounds and the minimum of all its upper bounds. 

Knowing the lower and upper bound of variables is enough to generate the loop nests. By taking the same steps over the redundancy map, boundaries can be determined, and the final result can be reconstructed. If there is a chain of computations, first all of them are done in the compressed space, and then the final result is reconstructed based on the output structure.

To generate code for a program that is written in the form of a sum of products, our code generator iterates over each multiplication, and generates loop nests per multiplication. In the deepest loop nest, the value of the result is updated by the value of the tensor expression appearing in the body of the rule.

\systemname{} hoists loop-invariant expressions outside the loops whenever possible. However, there is no data-layout optimization applied~\cite{chou2018format,schleich2022optimizing,kandemir1999linear}; an $n$ dimensional tensor is stored in an $n$ dimensional C++ array. Further optimizations such as multithreading and vectorization are also left for the future.

\examplepara{Vector Self-Outer-Product}
\begin{figure}
    \centering
    \begin{tabular}{|l|l}
        \begin{lstlisting}[language=C++, basicstyle=\scriptsize]
for(int i = 0; i < n; ++i) {
  auto y_i = y[i];
  auto x_i = x[i];
  for(int j = i; j < n; ++j) {
    y_i[j] = x_i * x[j];
  }
}
        \end{lstlisting} & 
        \begin{lstlisting}[language=C++, basicstyle=\scriptsize]
for(int i = 0; i < n; ++i) {
  int ip = j;
  auto y_i = y[i];
  auto y_ip = y[ip];
  for(int j = 0; j < i; ++j) {
    int jp = i;
    y_i[j] = y_ip[jp];
  }
}
        \end{lstlisting}
    \end{tabular}
    \caption{Code snippets for linear regression covariance matrix creation compressed computation (left) and result matrix reconstruction (right).}
    \label{fig:lrcode}
\end{figure}
The outer product of a vector with size $n$ by itself in \unifiedir{} is defined as follows.

\begin{centering} 
\begin{tabular}{r c l}
$y(i, j)$ & $:=$ & $x(i) * x(j)$ \\
$x_U(i)$ & $:=$ & $0 \leq i < n$ \\
$x_R(i, i')$ & $:=$ & $\emptyset$
\end{tabular}
\end{centering} 

\noindent
Based on Figure~\ref{fig:redundancymap}, output structure is:

\begin{tabular}{r c l}
$y_U(i, j)$ & $:=$ & $(0 \leq i < n) * (0 \leq j < n) * (i \leq j)$ \\ 
& $:=$ & $0 \leq i \leq j< n$  \\
$y_R(i, j, i', j')$ & $:=$ & $(0 \leq i < n) * (0 \leq j < n) * (i > j) * (i' = i) * (j' = j)$ \\
& $:=$ & $(0 \leq j < i < n) * (i' = i) * (j' = j)$
\end{tabular}

\noindent
The structure is fed to the code generator, and compressed computational loop nests are generated based on the unique set of $y$. Since all the elements in matrix $y$ should be accessible in the end, reconstruction code is generated based on the redundancy map of the output. Figure~\ref{fig:lrcode} shows the computation and reconstruction code snippets for this program.

\input{semantics}

\section{Experimental Results}
\label{sec:exp}
In this section, we experimentally evaluate \systemname{} by considering several tensor processing kernels over different tensor structures. We study the following questions:

\begin{itemize}
    \item How does leveraging the redundancy structure affect the run-time performance and in which cases \systemname{} performs better than the state-of-the-art dense tensor algebra frameworks?
    \item  How advantageous is symbolic sparsity over dynamic sparsity?
    \item How practical is \systemname{} in real-world applications? Is it worthwhile to perform the computation over a compressed tensor and then decompress the result in practice?
\end{itemize}

\subsection{Experimental Setup}
We use a server with operating system 18.04.5 LTS Ubuntu equipped with 10 cores 2.2 GHz Intel Xeon Silver 4210 CPU, and 220 GB of main memory. C++ code is compiled with GCC 7.5.0 using C++17 and -O3, -pthread, -mavx2, -ffast-math, and -ftree-vectorize flags. All experiments are run on a single thread, and the average run-time of five runs is reported. 
As the competitors, we use the latest version of TACO\footnote{https://github.com/tensor-compiler/taco/tree/2b8ece4c230a5f}~\cite{kjolstad:2017:taco}, Numpy 1.23.4~\cite{harris2020array}, PyTorch 1.12.1~\cite{Paszke_PyTorch_An_Imperative_2019}, and TensorFlow 2.10.0~\cite{tensorflow2015-whitepaper} with and without the XLA backend.

\begin{figure}[t]
\setlength\tabcolsep{.5pt}
    \centering
    \begin{tabular}{ccc}
         \includegraphics[width=0.33\columnwidth]{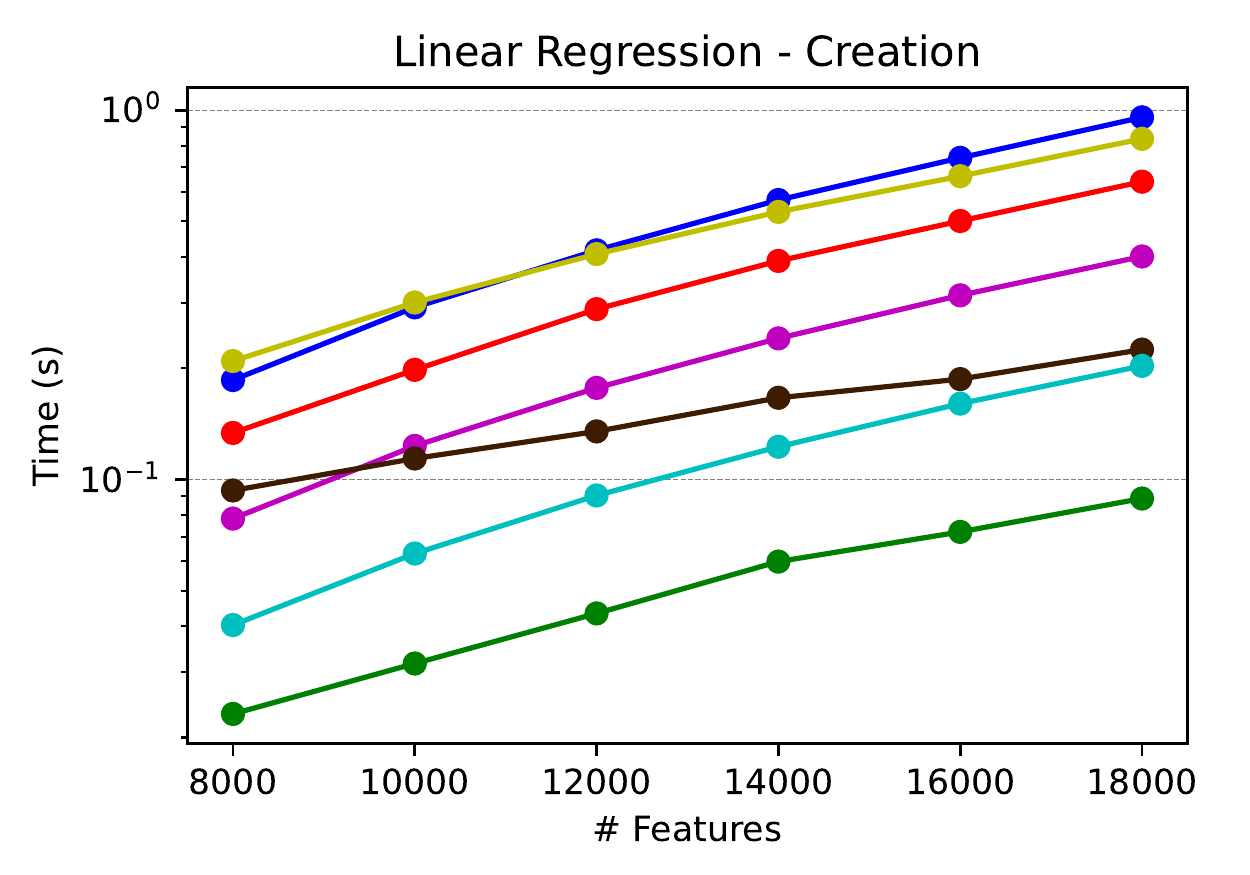} & 
         \includegraphics[width=0.33\linewidth]{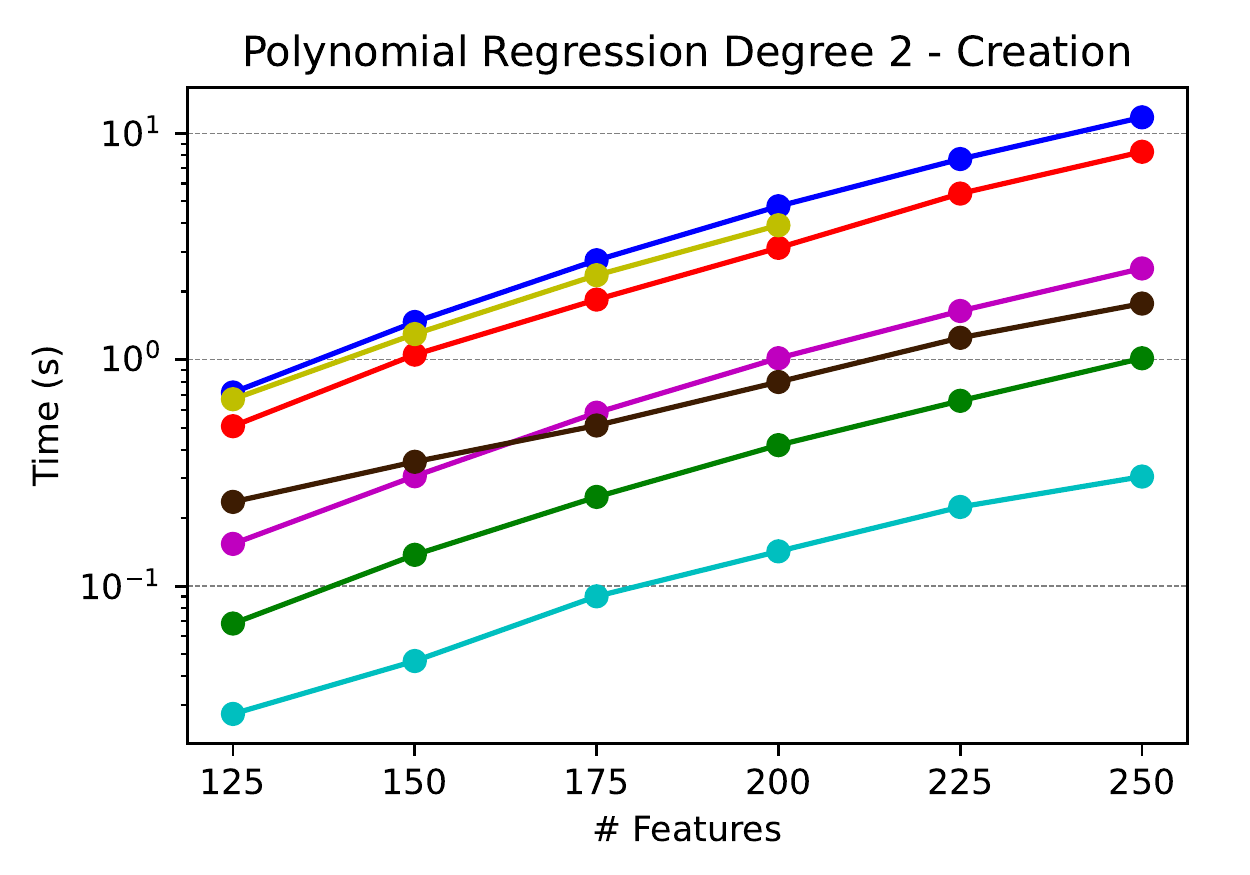} & 
         \includegraphics[width=0.33\linewidth]{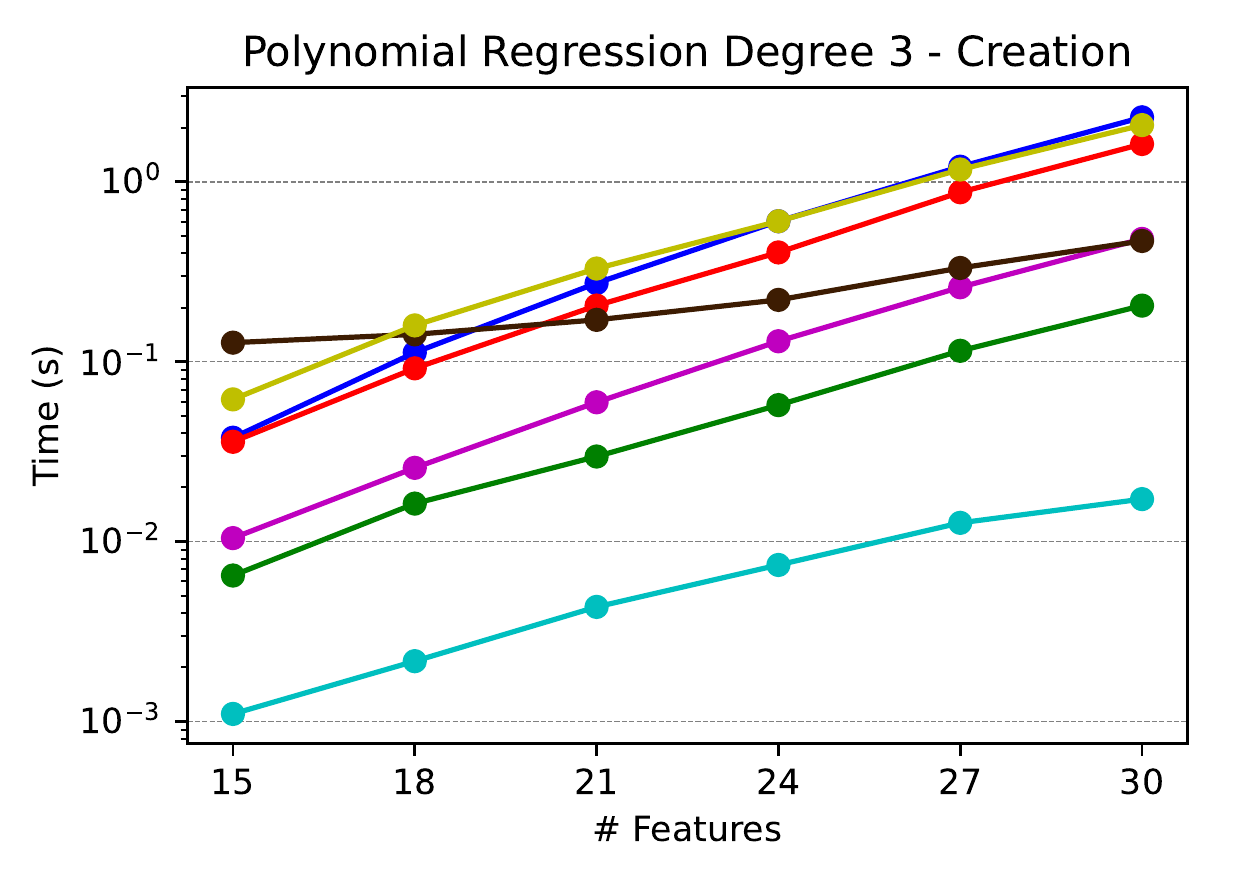} \\
         \includegraphics[width=0.33\linewidth]{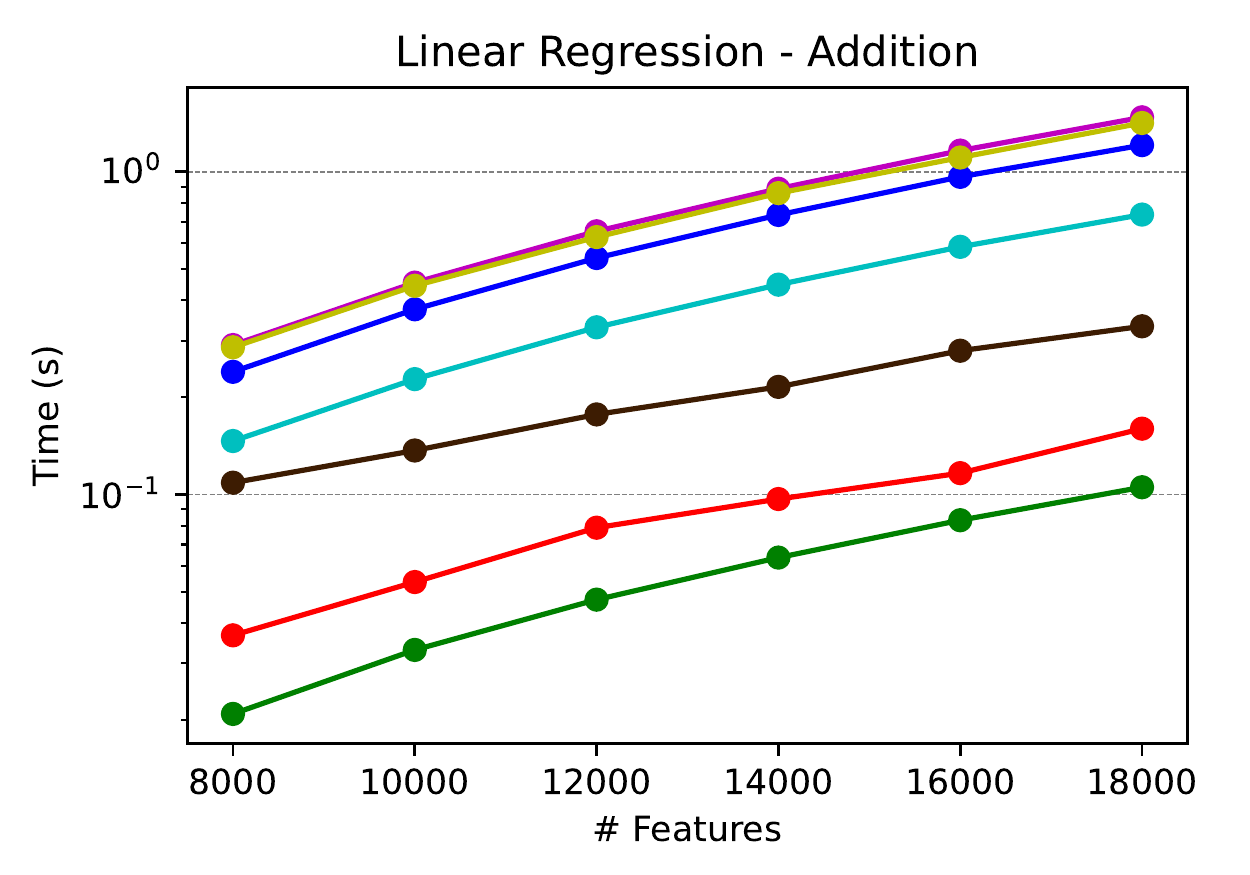} & \includegraphics[width=0.33\linewidth]{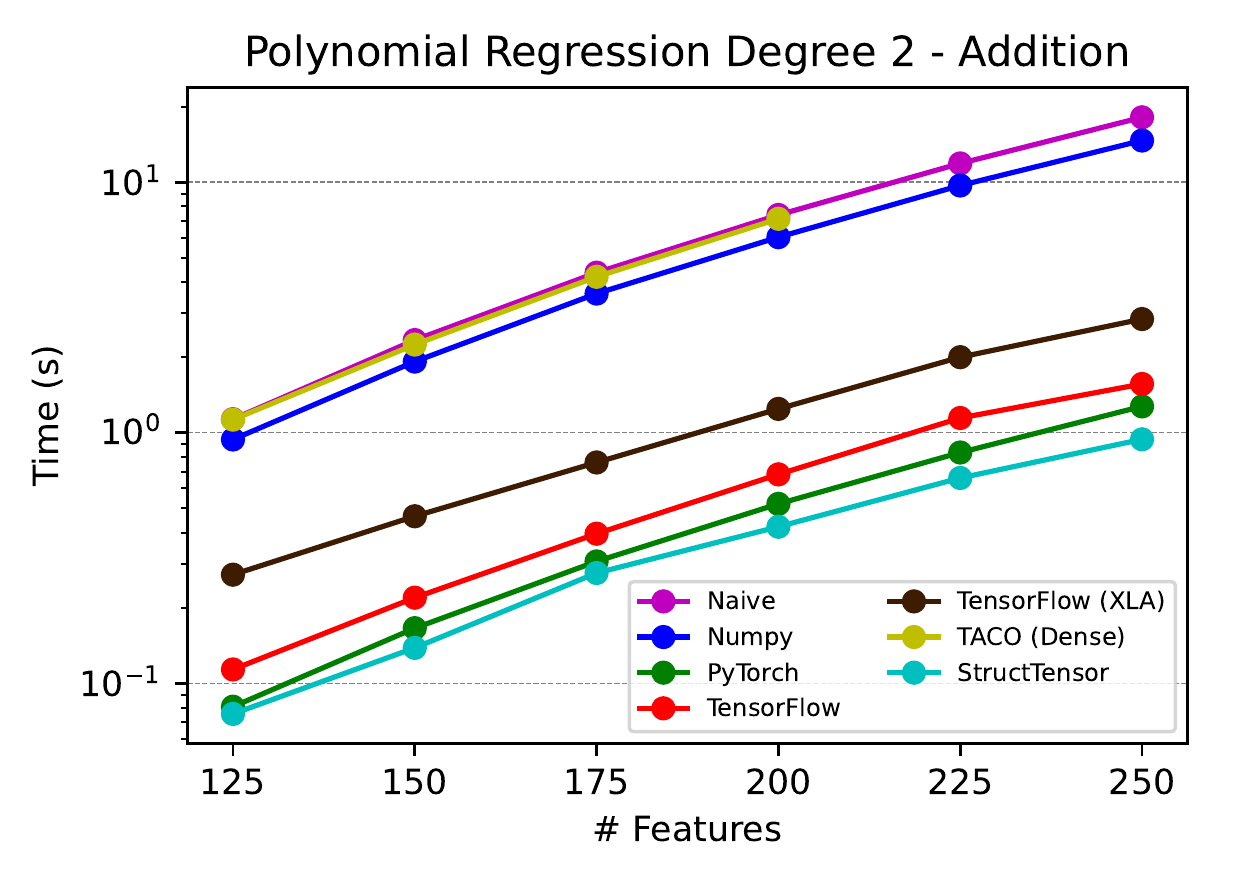} & \includegraphics[width=0.33\linewidth]{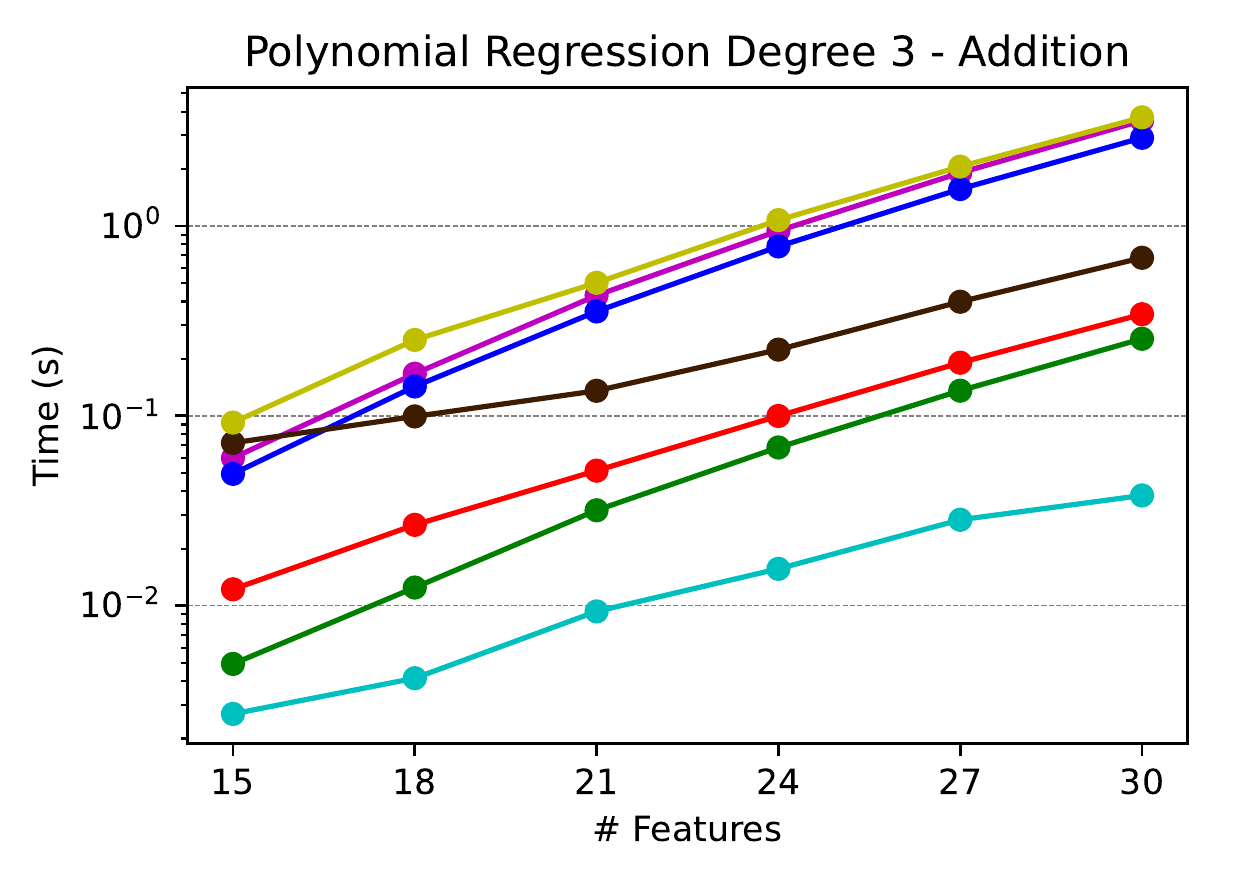}
    \end{tabular}
    \caption{The run time comparison of the covariance matrix creation and addition for linear regression and polynomial regression degree-2 and degree-3. For linear regression dense frameworks show better performance, as the redundancy structure does not pay off. For models with higher redundancy, \systemname{} significantly outperforms the competitors thanks to avoiding redundant computation. }
    \label{fig:covar}
\end{figure}

\subsection{Redundancy Structure}
\begin{table}
\scriptsize
\begin{tabular}{|c|c|c|c|}
\hline
\textbf{Kernel} & \textbf{Structure of $B$} & \textbf{$B_U$ in \unifiedir{}} & TACO (Smart)\\
\hline
\textbf{TTM} & Diagonal (plane) & $(0 \leq i < n_i) * (i = j) * (0 \leq l < n_l)$ & $(D, S, D) * (D, D) \rightarrow (D, S, D)$ \\
$A(i, j, k) :=$ & Fixed $j$ & $(0 \leq i < n_i) * (j = J) * (0 \leq l < n_l)$ & $(D, S, D) * (D, D) \rightarrow (D, S, D)$\\
$B(i, j, l) * C(k, l)$ & Upper half cube & $(0 \leq i < n_i) * (i \leq j < n_j) * (0 \leq l < n_l)$ & $(D, S, D) * (D, D) \rightarrow (D, S, D)$\\
\hline
\textbf{THP}  & Diagonal (plane) & $(0 \leq i < n_i) * (i = j) * (0 \leq l < n_l)$ & $(D, S, D) * (D, D, D) \rightarrow (D, S, D)$ \\
$A(i, j, k) :=$ & Fixed $i$ & $(i = I) * (0 \leq j < n_j) * (0 \leq l < n_l)$ & $(S, D, D) * (D, D, D) \rightarrow (S, D, D)$\\
$B(i, j, k) * C(i, j, k)$ & Fixed $j$ & $(0 \leq i < n_i) * (j = J) * (0 \leq l < n_l)$ & $(D, S, D) * (D, D, D) \rightarrow (D, S, D)$ \\
\hline
\textbf{MTTKRP} & Fixed $i, j$ & $(i = I) * (j = J) * (0 \leq l < n_l)$ & $(S, D, D) * (D, D) * (D, S) \rightarrow (S, D)$ \\
$A(i, j) := B(i, k, l)$ & Fixed $i$ & $(i = I) * (0 \leq j < n_j) * (0 \leq l < n_l)$ & $(S, D, D) * (D, D) * (D, D) \rightarrow (S, D)$\\
$ * $ $ C(k, j) * D(l, j)$ & Fixed $j$ & $(0 \leq i < n_i) * (j = J) * (0 \leq l < n_l)$ & $(D, D, D) * (D, D) * (D, S) \rightarrow (D, S)$\\
\hline
\end{tabular}
\caption{List of tensor kernels for \systemname{} evaluation. TACO allows specifying if each dimension is Sparse ($S$) or Dense ($D$). TACO (Smart) specifies the most efficient storage format.}
\label{tbl:kernels}
\end{table}

\begin{figure}
\setlength\tabcolsep{.5pt}
    \centering
    \begin{tabular}{ccc}
         \includegraphics[width=0.33\linewidth]{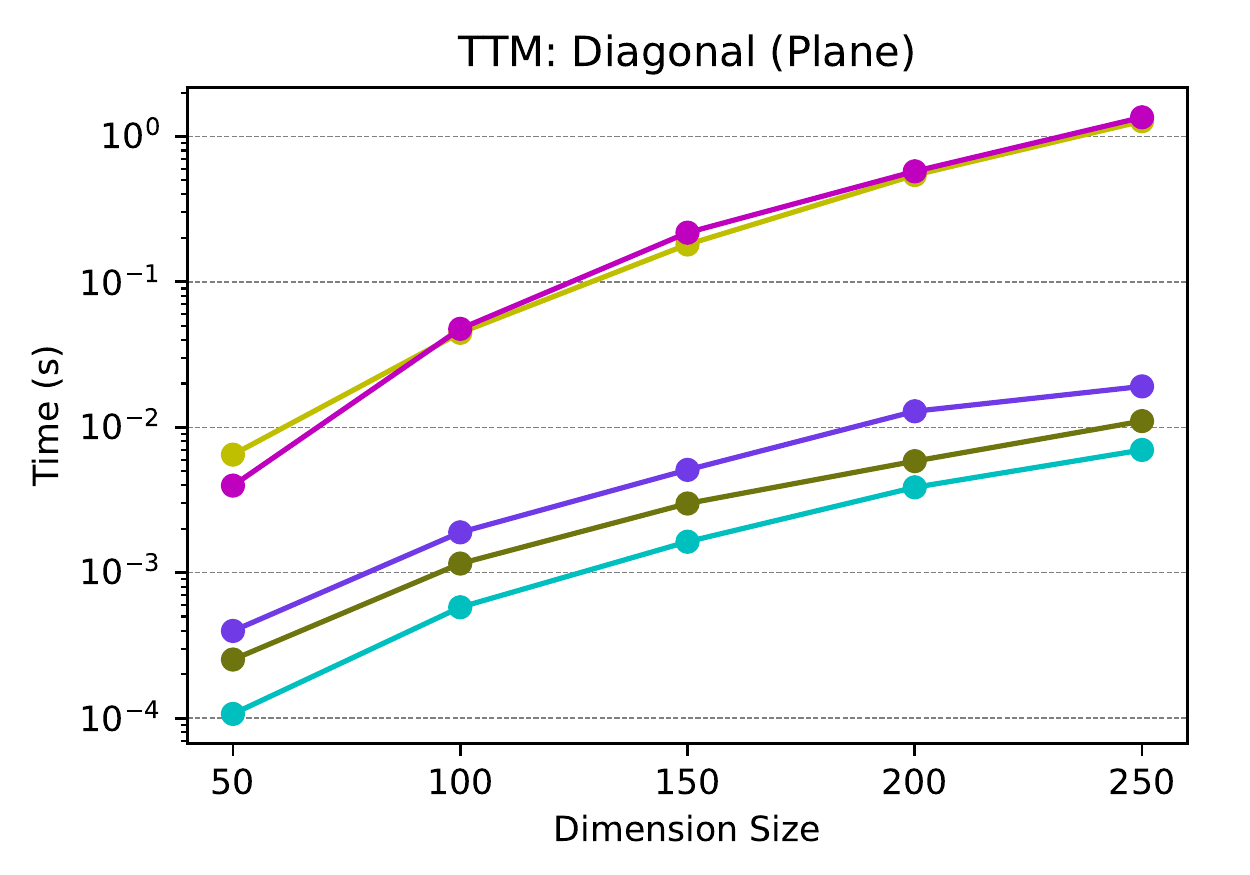} & \includegraphics[width=0.33\linewidth]{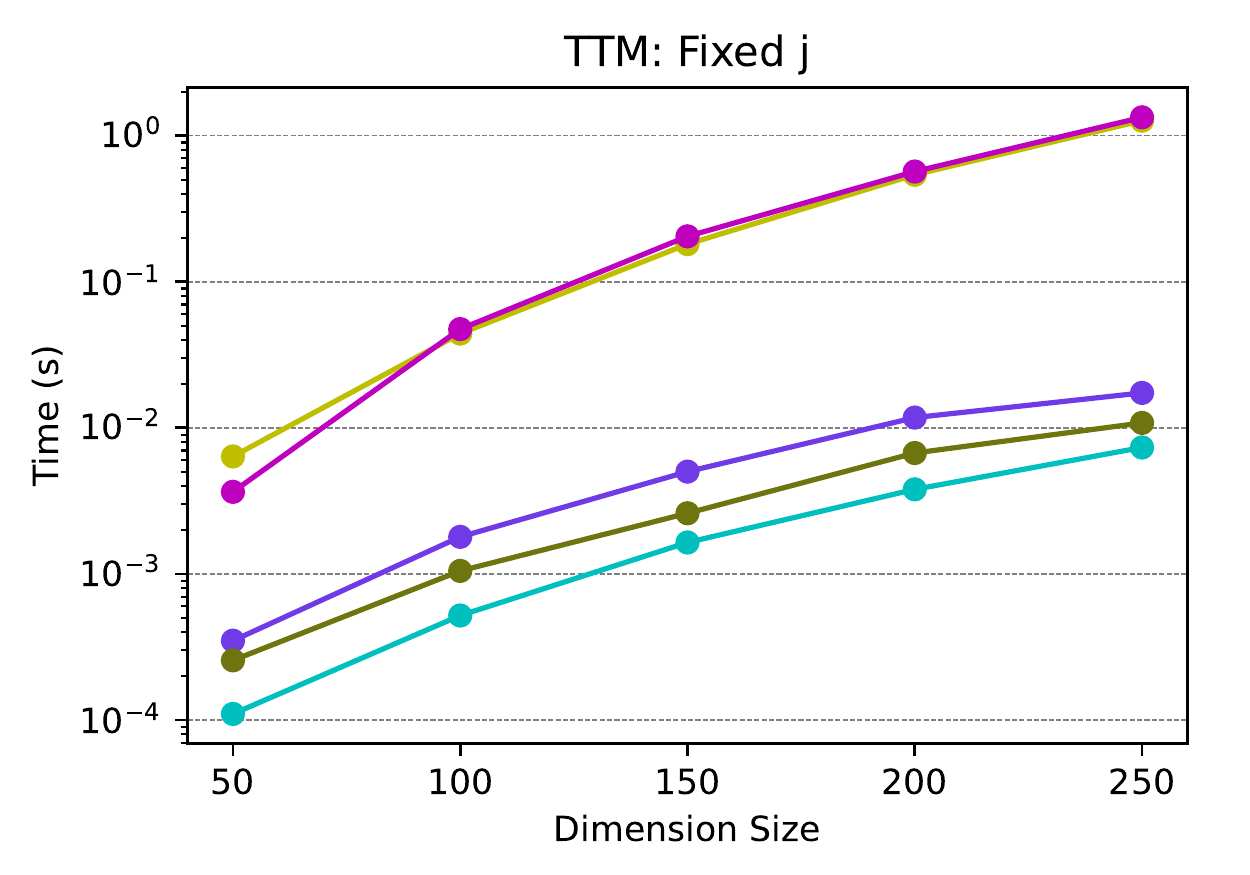} & \includegraphics[width=0.33\linewidth]{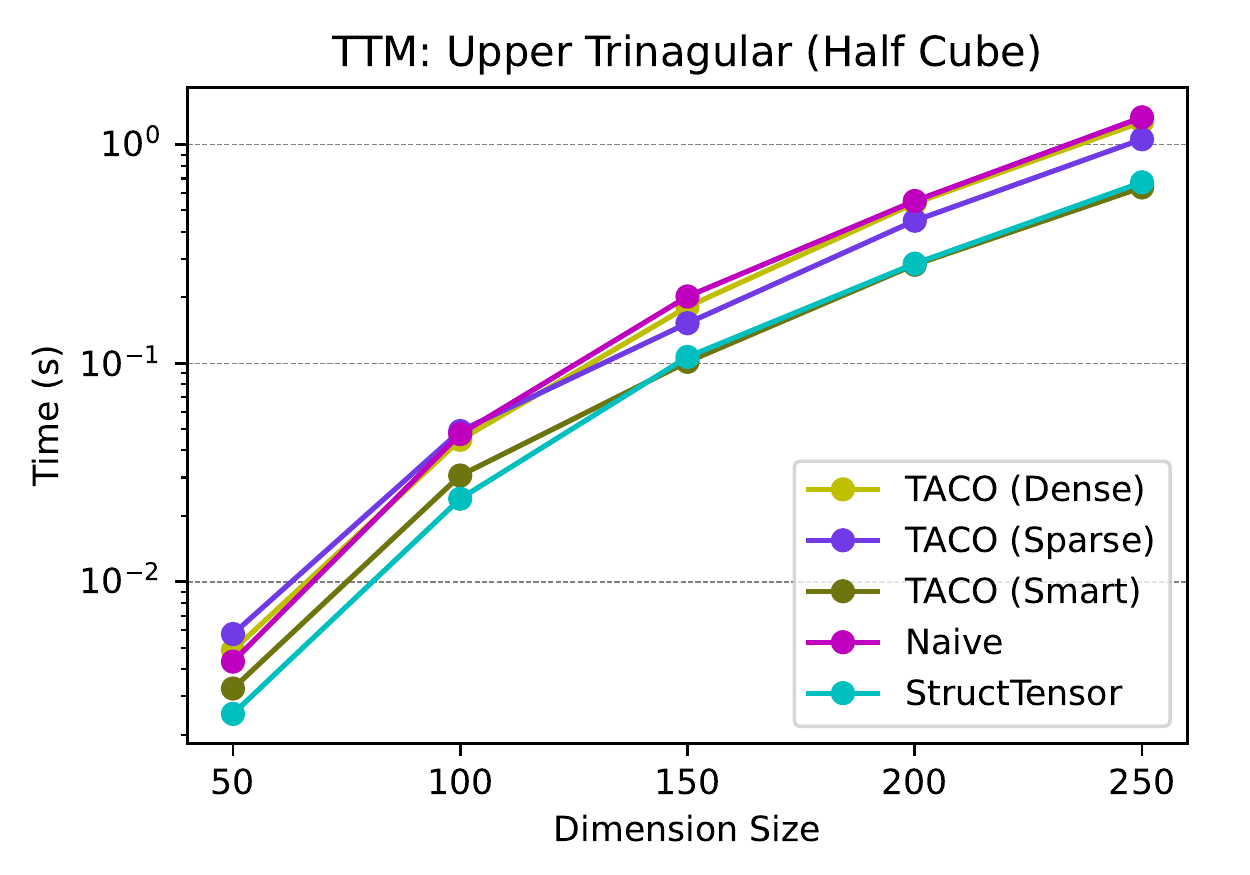} \\
         \includegraphics[width=0.33\linewidth]{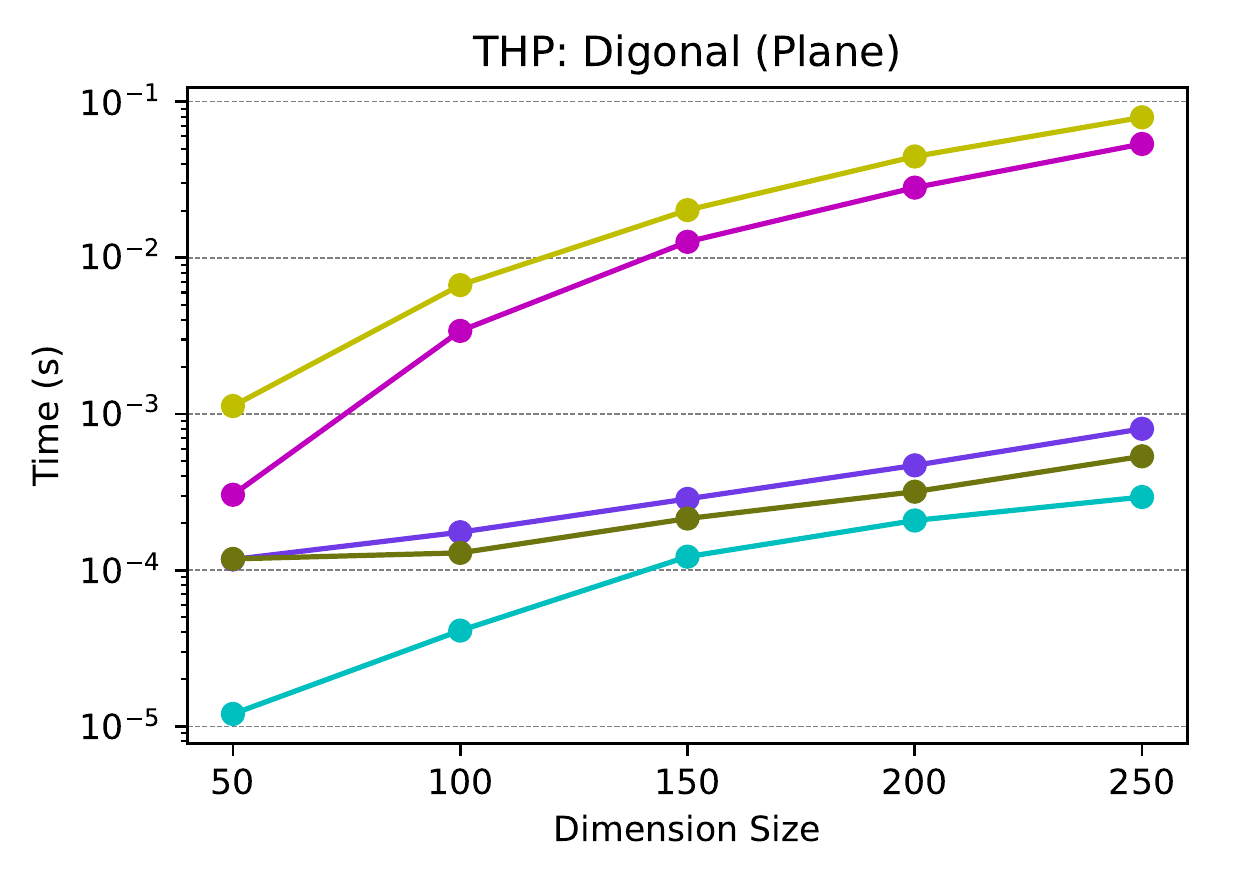} & \includegraphics[width=0.33\linewidth]{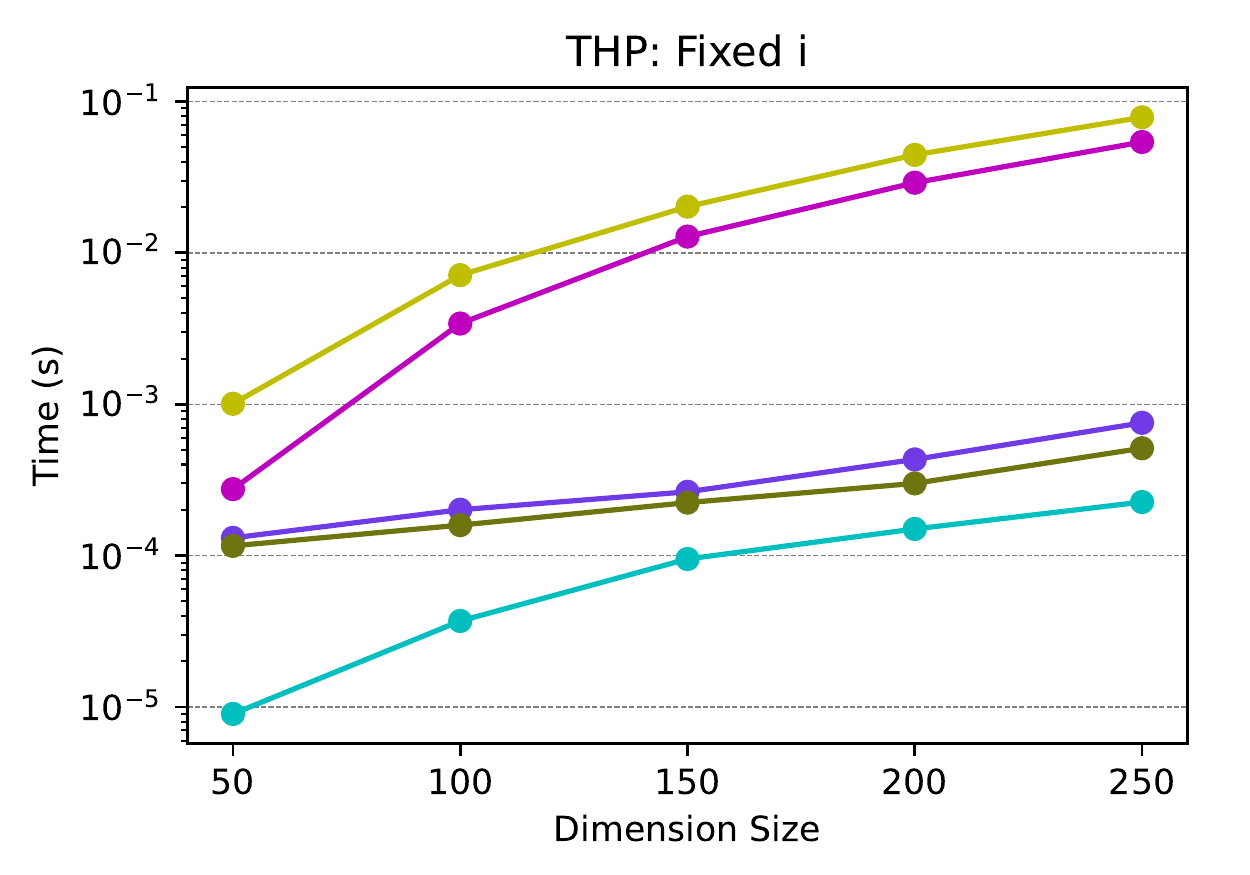} & \includegraphics[width=0.33\linewidth]{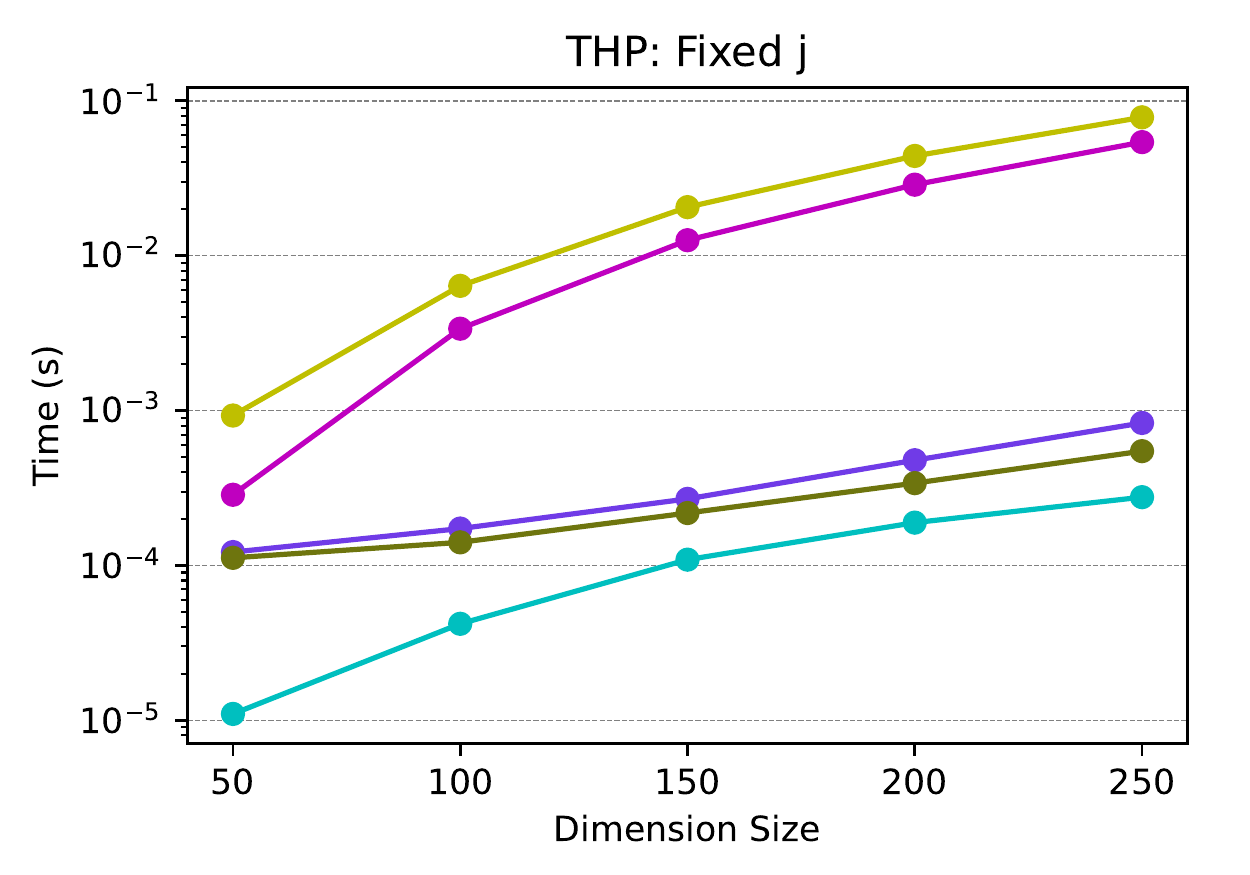} \\
         \includegraphics[width=0.33\linewidth]{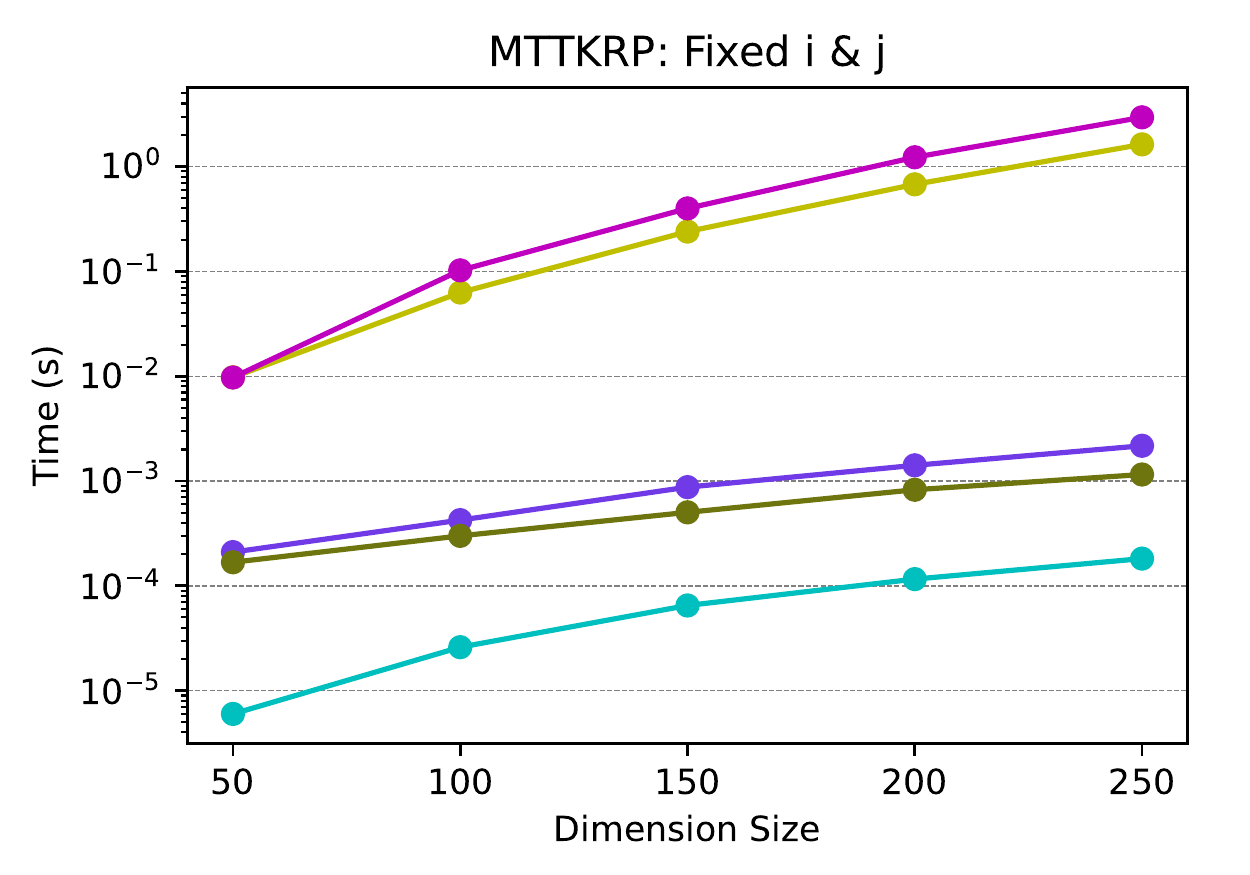} & \includegraphics[width=0.33\linewidth]{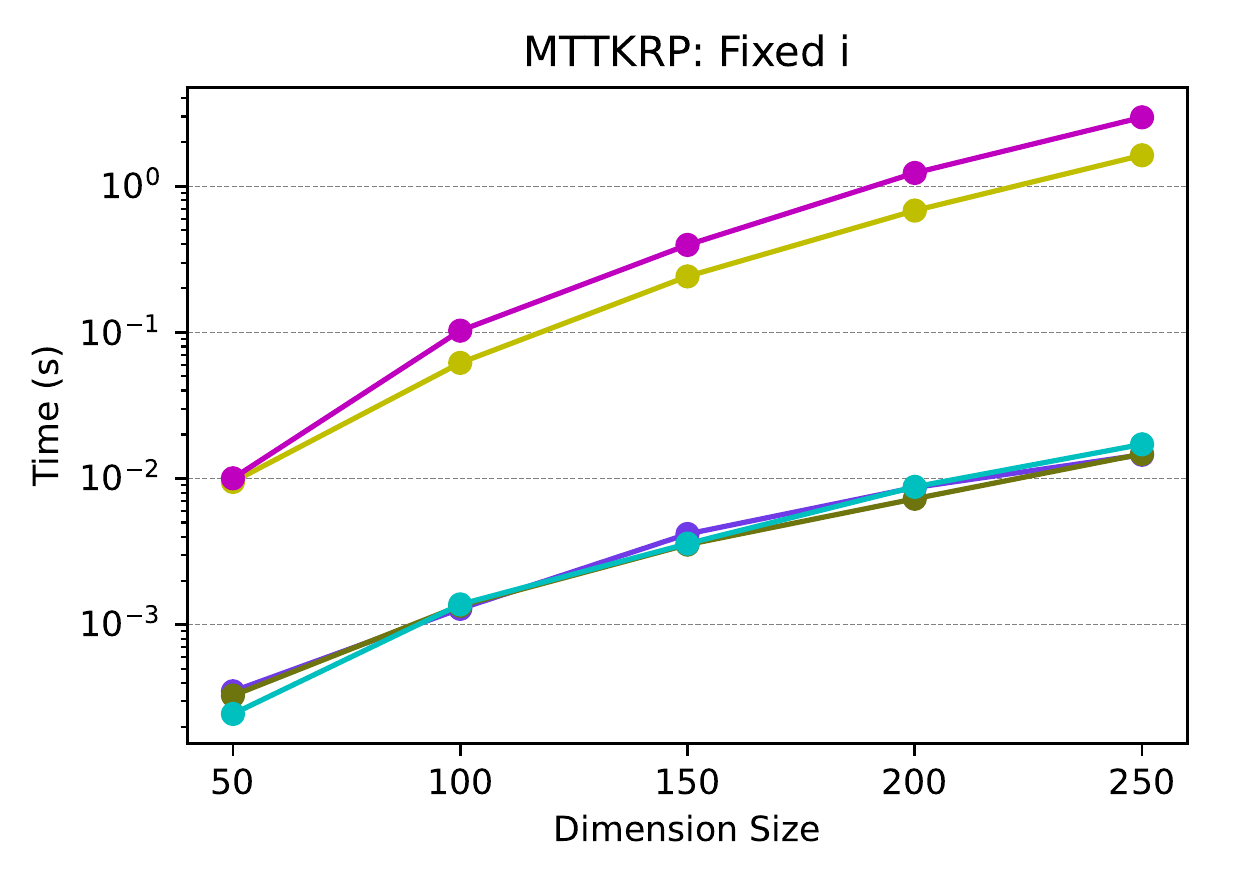} & 
         \includegraphics[width=0.33\linewidth]{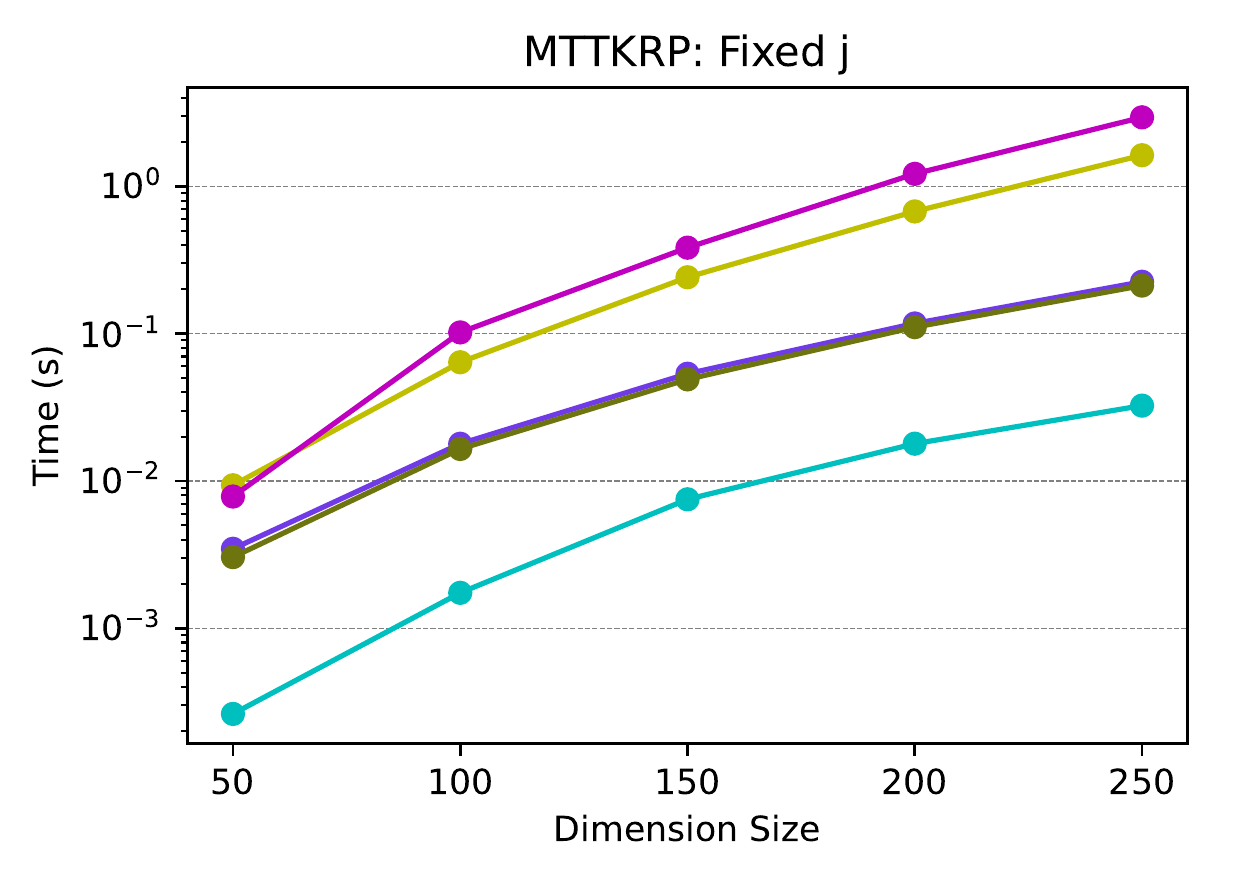}
    \end{tabular} 
    \caption{Tensor operation kernels run time comparison using different input and output structures. The na\"ive version of \systemname{} and the dense format of TACO show similar performance. In most cases, the symbolic sparsity employed in \systemname{} shows significantly better performance in comparison with the best sparse format used in TACO.}
    \label{fig:indkernels}
\end{figure}

We consider six structured matrix kernels that correspond to the task of creating and addition of covariance matrices for three machine learning models: linear regression, polynomial regression degree-2, and polynomial regression degree-3.
As all the matrices only involve redundancy structure, we need to use the dense representation in all systems.
As the competitors, we consider TACO (fully dense format), PyTorch, TensorFlow (w/ and w/o the XLA backend), NumPy, and a na\"ive implementation without leveraging the redundancy structure. 

Figure~\ref{fig:covar} shows the run time comparison. The missing numbers mean that the process was killed due to a segmentation fault. For example, TACO can only handle the polynomial regression degree-2 up to 200 features because it gets killed for more than that. For the kernels with more redundancy (i.e., polynomial regression degree-2 and degree-3), \systemname{} outperforms all the competitors from one to two orders of magnitude. However, in linear regression \systemname{} only removes half of the computation and because of a lack of data-layout optimizations, loop tiling, and vectorization, performs worse than dense tensor frameworks such as PyTorch and TensorFlow. 

\subsection{Sparsity Structure}
To evaluate the effectiveness of \systemname{} for the sparsity structure, we consider three tensor kernels: Tensor Times Matrix (TTM), Tensor Hadamard Product (THP), and Matricized Tensor Times Khatri-Rao Product (MTTKRP). \systemname{} is evaluated against TACO with three different formats: fully dense, fully sparse, and smart. In the smart version, we use the most efficient sparse format for TACO based on the sparsity of input and output tensors at each dimension. 
Table~\ref{tbl:kernels} shows the definition of these kernels, the different input structures we considered, their representation in \unifiedir{}, and the data format at each dimension for TACO (Smart).\footnote{Note that the output data format in MTTKRP for both dimensions in TACO (Smart) for the case of fixed $i, j$ and all cases of TACO (Sparse) should be $(S, S)$. However, because of a bug in TACO for sparse outputs (https://github.com/tensor-compiler/taco/issues/518), we had to use $(D, S)$ or $(S, D)$ instead.} 
For \systemname{} we consider an additional na\"ive version that does not leverage the symbolic structure.

Figure~\ref{fig:indkernels} represents the run time of each implementation on each kernel.
In all kernels i) TACO dense is performing similarly to the na\"ive implementation, ii) TACO smart is outperforming other data format selections for TACO as expected, and iii) \systemname{} outperforms (in 7 out of 9 experiments) or performs on par with TACO smart (despite having a better data layout and generating a more optimized code).

\begin{figure}
    \centering
    \begin{tabular}{cc}
    \includegraphics[width=0.45\linewidth]{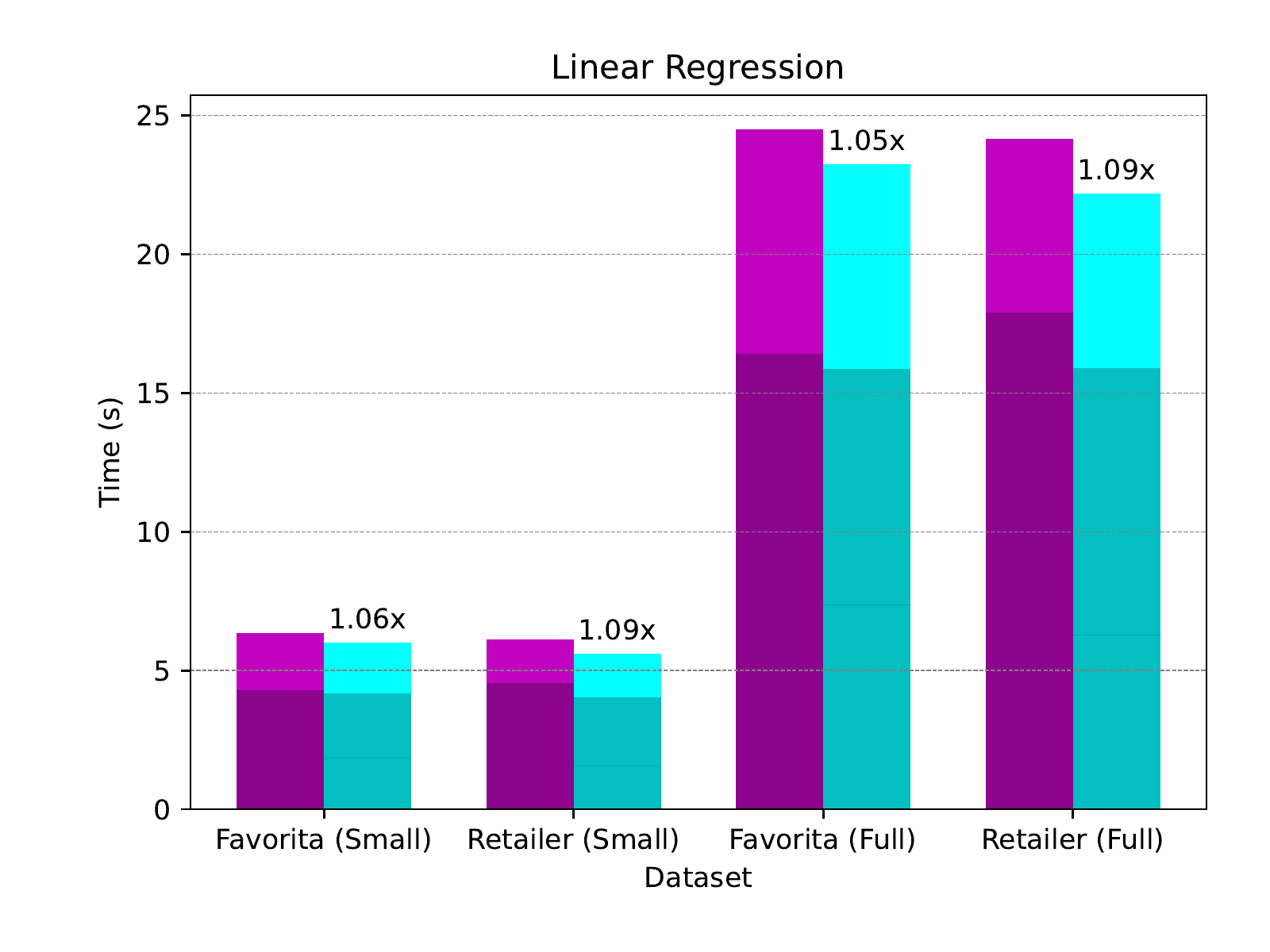} & \includegraphics[width=0.45\linewidth]{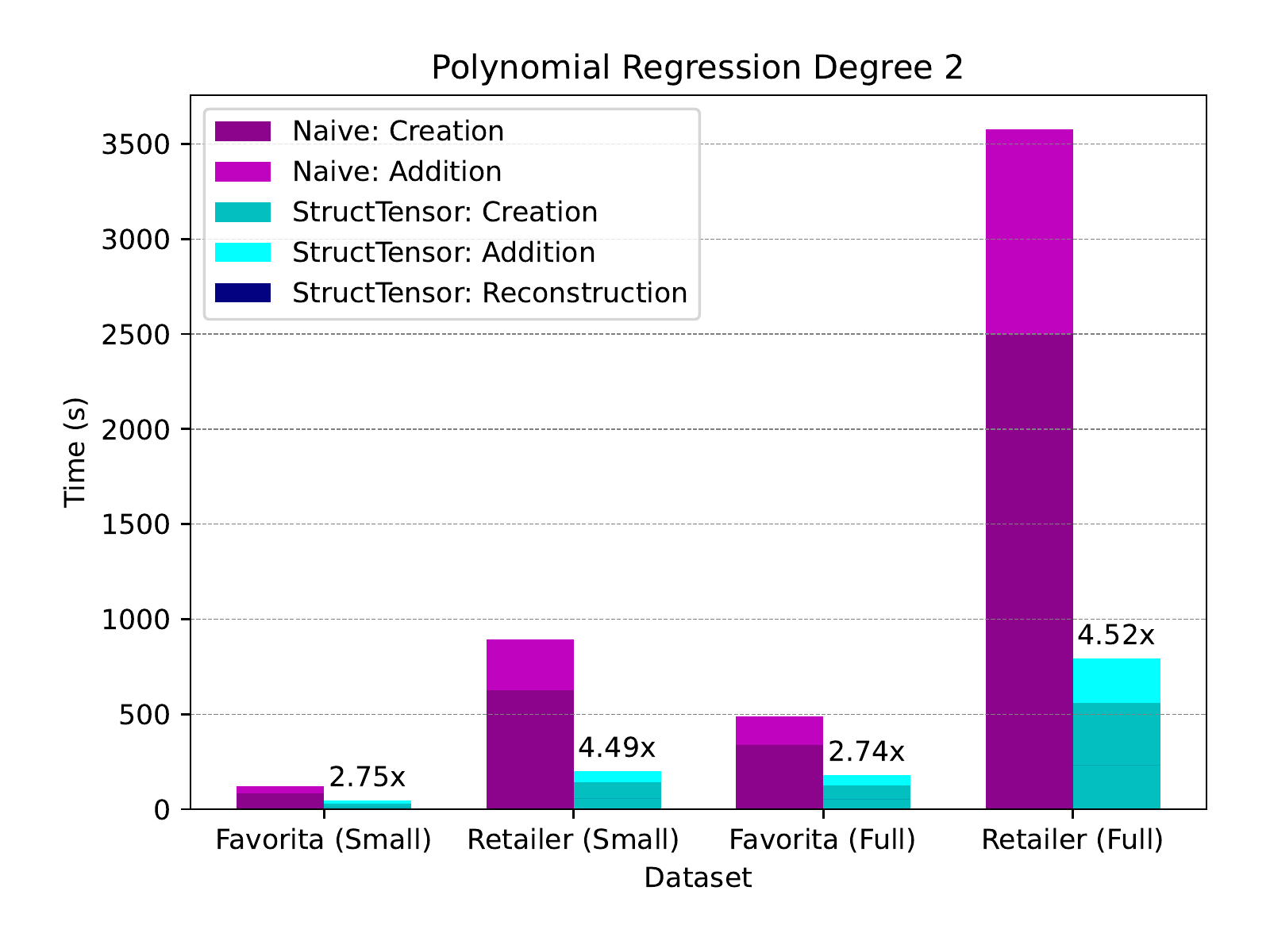}
    \end{tabular}
    \caption{Run time of in-database machine learning over Favorita and Retailer datasets. Here, we compute the covariance matrix of linear and polynomial regression degree-2 over the join of multiple relations.}
    \label{fig:e2e}
\end{figure}

\subsection{End-to-end Experiments}
\label{sec:exp:e2e}
We evaluated \systemname{} on an in-database machine learning task. We used \systemname{} to create a covariance matrix for linear and polynomial regression degree-2. We consider the following real-world datasets: Retailer~\cite{khamis2020learning} with 13 continuous features and Favorita~\cite{favorita} with 6 continuous features. Both of them have a small (consisting of 25\% of the data elements) and full version. \systemname{} is compared with a na\"ive implementation that does not leverage the redundancy structure. 

We use the idea of a semi-ring covariance matrix data-structure~\cite{10.1145/3183713.3183758,shaikhha2022functional} for both implementations. 
This data structure pushes joins after aggregates, which heavily improves both memory consumption and run time. In a semi-ring covariance matrix, different degrees of interaction are decoupled and stored separately. This means that for polynomial regression degree-2 (cf. Section~\ref{sec:overview}), we store degree-2 ($M1$), degree-3 ($M2$), and degree-4 ($M3$) interactions separately.
Furthermore, degree-3 interactions are calculated twice in both ($\mathbf{f} \otimes \vectorize{\mathbf{f} \otimes \mathbf{f}}$) and ($\vectorize{\mathbf{f} \otimes \mathbf{f}} \otimes \mathbf{f}$). 
As the na\"ive version does not use the structure information, it cannot detect such coarse-grained redundancy information.

Figure~\ref{fig:e2e} represents the run time of \systemname{} in comparison with the na\"ive version.
\systemname{} produces a structure-aware code that reduces the computations by avoiding redundant computation. Hence, there should be a reconstruction phase that rebuilds the final result and put all elements that have not been computed in their corresponding positions based on the redundancy map. However, the reconstruction time is negligible in comparison with the rest of the computation, as it can be seen in Figure~\ref{fig:e2e}.

\input{related}

\section{Conclusion \& Outlook}
In this paper, we presented \systemname{}, a compiler for structured tensor algebra.
We considered two classes of structures: (1) sparsity patterns, and (2) redundancy structures.
We proposed \unifiedir{}, a unified IR that is expressive enough for tensor computation and captures both forms of structures. 
We have shown the soundness of transformations and inference rules.
Finally, the experimental results show that \systemname{} outperforms the state-of-the-art tensor processing libraries. 

For the future, we see two clear directions for improvement. 
First, \systemname{} mainly focuses on structure-specific optimizations, and apart from code motion, does not do anything smart for loop optimizations. 
It would be interesting to use polyhedral-based optimizations (e.g., using CLooG~\cite{bastoul2004code}) to generate efficient loop nests.
Second, for storing tensors, we mainly use $n-d$ arrays, and even the compressed tensors still allocate the memory for the entire uncompressed tensor. 
Using better layouts can significantly reduce the memory pressure, and would possibly improve data locality.

\bibliography{refs}

\end{document}

%% file: macros.tex
\newcommand{\amirsh}[1]{{\color{red}Amir: #1}}
\newcommand{\note}[1]{{\color{blue}Note: #1}}
\newcommand{\mh}[1]{{\color{orange}Mathieu: #1}}
\newcommand{\systemname}{{\sc StructTensor}}
\newcommand{\unifiedir}{{\sc STUR}}
\newcommand{\smartpara}[1]{\noindent \textbf{#1.}}
\newcommand{\lemexp}[1]{\noindent \normalfont {#1.}}
\newcommand{\translatebegin}{$\llbracket$}
\newcommand{\translateend}{$\rrbracket$}
\newcommand{\translate}[1]{\translatebegin{}\text{#1}\translateend{}}
\newcommand{\metavar}[1]{$#1$}
\newcommand{\metavars}[1]{$\MakeLowercase{#1}$}
\newcommand{\binop}{\diamond}
\newcommand{\domain}[1]{FV(#1)}

\newcommand{\matrixmult}{\cdot}
\newcommand{\tab}{$~~~$}
\newcommand{\argvec}[1]{\bm{#1}}
\newcommand\sbullet[1][.75]{\mathbin{\vcenter{\hbox{\scalebox{#1}{$\bullet$}}}}}
\newcommand{\vectorize}[1]{vec(#1)}
\newcommand{\hcat}{||}
\newcommand{\vcat}{//}

\newcommand{\TT}{\mathcal{T}}
\newcommand{\RR}{\mathbb{R}}
\newcommand{\II}{\mathbb{I}}
\newcommand{\BB}{\mathbb{B}}
\newcommand{\forget}[1]{\lfloor #1 \rfloor}

\definecolor{colg}{rgb}{0.1,0.7,0.1}
\definecolor{colr}{rgb}{0.7,0.1,0.1}
\definecolor{colb}{rgb}{0.1,0.1,0.7}
\definecolor{colbb}{rgb}{0.1,0.1,0.6}
\newcommand{\supfull}{{\color{colg}{\CIRCLE}}}
\newcommand{\suphalf}{{\color{colb}{\LEFTcircle}}}
\newcommand{\supnone}{{\color{colr}{\Circle}}}

\newcommand{\examplepara}[1]{\smartpara{Example - #1}}

%% file: figures/grammar-la.tex
\begin{figure}
\begin{tabular}{c c}
\begin{tabular}{|c|l|}
\multicolumn{2}{c}{$e$ $::=$ $e \matrixmult e$ $\mid$ $e \odot e$ $\mid$ $e \otimes e$ $\mid$ $e + e$ $\mid$ $e \oplus e$ $\mid$  $e^T$} \\
\multicolumn{2}{c}{}\\
\hline
$\mathcal{G}$    & General dense matrix \\ 
$\mathcal{S}$   & Symmetric matrix  \\
 $\mathcal{D}$   & Diagonal matrix  \\
 $\mathcal{R}_n$    & Matrix of non-zeros only at $n^{th}$ row \\ 
  $\mathcal{C}_n$    & Matrix of non-zeros only at $n^{th}$ column \\ 
$\mathcal{H}_{n,m}$    & Singular matrix at index $n,m$ \\ 
$\mathcal{Z}$    & All zeros matrix \\ 
  \hline
\end{tabular} &
\begin{tabular}{c}
\begin{tabular}{c}
$e$: $\mathcal{C}_{n}$
\\ \hline
$e \matrixmult e^T$: $\mathcal{S}$
\end{tabular} \hspace{0.5cm}
\begin{tabular}{c}
$e_1$: $\mathcal{C}_{n}$ \tab $e_2$: $\mathcal{R}_{m}$ \tab $n \neq m$
\\ \hline
$e_1 \matrixmult e_2$: $\mathcal{Z}$
\end{tabular} 
\\
\begin{tabular}{c}
$e_1$: $\mathcal{C}_{n}$ \tab $e_2$: $\mathcal{R}_{n}$
\\ \hline
$e_1 \matrixmult e_2$: $\mathcal{G}$
\end{tabular} \hspace{0.5cm} 
\begin{tabular}{c}
$e_1$: $\mathcal{R}_{n}$ \tab $e_2$: $\mathcal{C}_{m}$
\\ \hline
$e_1 \times e_2$: $\mathcal{H}_{n, m}$
\end{tabular}
\\
\begin{tabular}{c}
$e_1$: $\mathcal{R}_{n}$ \tab $e_2$: $\mathcal{R}_{m}$
\\ \hline
$e_1 \matrixmult e_2$: $\mathcal{R}_{n}$ 
\end{tabular} \hspace{0.5cm} 
\begin{tabular}{c}
$e_1$: $\mathcal{C}_{n}$ \tab $e_2$: $\mathcal{C}_{m}$
\\ \hline
$e_1 \matrixmult e_2$: $\mathcal{C}_{m}$
\end{tabular} 
\\

\begin{tabular}{c}
$e_1$: $\mathcal{D}$ \tab $e_2$: $\mathcal{D}$
\\ \hline
$e_1 \odot e_2$: $\mathcal{D}$
\end{tabular} 
\hspace{0.5cm} 
\begin{tabular}{c}
$e_1$: $\mathcal{D}$ \tab $e_2$: $\mathcal{D}$
\\ \hline
$e_1 \otimes e_2$: $\mathcal{D}$
\end{tabular}

\\

\begin{tabular}{c}
$e_1$: $\mathcal{D}$ \tab $e_2$: $\mathcal{D}$
\\ \hline
$e_1 \matrixmult e_2$: $\mathcal{D}$
\end{tabular} \hspace{0.5cm}
\begin{tabular}{c}
$e_1$: $\mathcal{D}$ \tab $e_2$: $\mathcal{D}$
\\ \hline
$e_1 \oplus e_2$: $\mathcal{D}$
\end{tabular} \\

\end{tabular}
\end{tabular}
\caption{The grammar for structured linear algebra operations and a subset of structure inference rules.}
\label{fig:structla}
\end{figure}

%% file: figures/grammar-core.tex
\begin{figure}
    \centering
    \begin{tabular}{|r r c l|l|}
      \hline
        Program & \metavars{P} & $::=$ & \metavars{R} $\mid$ \metavars{R}; \metavars{P} &  List of rules. \\ \hline
        Rule & \metavars{R} & $::=$ & \metavar{A} $:=$ \metavar{B} & Head (access) and body (Sum of Products). \\ \hline
        Body & \metavar{B} & $::=$ & \metavars{F} $\mid$ \metavars{F} $+$ \metavar{B} 
        & Sum of factor products.
        \\ \hline
        Factor & \metavars{F} & $::=$ & \metavars{E} $\mid$ \metavars{E} $*$ \metavars{F}
        & Product of expressions.
        \\ \hline
        Expression & \metavars{E} & $::=$ & \metavars{X} $\theta$ \metavars{I} $\mid$ \metavars{I} $\theta$ \metavars{X} $\mid$ \metavar{A} 
        & Comparison ($\theta$) or access.
        \\ \hline
        
        Index & \metavars{I} & $::=$ & \metavars{X} $\mid$ \metavars{C} $\mid$  \metavars{I} $\binop$ \metavars{I}
        & Variable, constant, or arithmetic ($\binop$) over indices.\\ \hline
        Access & \metavar{A} & $::=$ & \metavar{T}($\overline{\text{\metavars{X}}}$) $\mid$
        \metavar{T}$_C$($\overline{\text{\metavars{X}}}$) 
        $\mid$ & Tensor, compressed tensor,  \\
        & & &
        \metavar{T}$_U$($\overline{\text{\metavars{X}}}$) 
        $\mid$
        \metavar{T}$_R$($\overline{\text{\metavars{X}}}$) & unique set, and redundancy map access.
        \\ 
    \hline
    \end{tabular}
    \caption{Grammar of \unifiedir{}. The meta variables \metavars{X} and \metavar{T} range over the name of index variables and tensors, respectively.}
    \label{fig:grammar}
\end{figure}

%% file: figures/freevars.tex
\begin{figure}
\setlength\tabcolsep{2.5pt}
\begin{tabular}{c c c}
\begin{tabular}{c}
\begin{tabular}{c}
$T(x_1, \ldots, x_n) := e$
\\ \hline
$ \domain{T(x_1, \ldots, x_n)}  \subseteq \domain{e}$ 
\end{tabular} \\
$ \domain{T(x_1, \ldots, x_n)} =   \{x_1, \ldots, x_n\}$ 
\end{tabular} &
\begin{tabular}{c}
$ \domain{e1 + e2} =   \domain{e1} \cap \domain{e2}$ \\
$ \domain{e1 * e2} =   \domain{e1} \cup \domain{e2}$ \\
$ \domain{t1 \binop t2} = \domain{t1} \cup \domain{t2}$
\end{tabular} &
\begin{tabular}{c}
$ \domain{x \theta t} =  \{ x \} \cup \domain{t}$ \\
$ \domain{x} =  \{ x \}$ \\
$ \domain{c} = \emptyset$ 
\end{tabular}
\end{tabular}

\caption{Free variable rules.}
\label{fig:freevars}
\end{figure}

%% file: figures/la-in-stur.tex
\begin{figure}

\setlength\tabcolsep{3.5pt}
\begin{tabular}{c c}
\begin{tabular}{r c l}
\translate{$e^T$}$(i, j)$ & $:=$ & \translate{$e$}$(j, i)$ \\
\translate{$e_1 + e_2$}$(i, j)$ & $:=$ & \translate{$e_1$}$(i, j)$ $+$ \translate{$e_2$}$(i, j)$ \\
\translate{$e_1 \matrixmult e_2$}$(i, j)$ & $:=$ & \translate{$e_1$}$(i, k)$ $*$ \translate{$e_2$}$(k, j)$ \\
\translate{$e_1 \odot e_2$}$(i, j)$ & $:=$ & \translate{$e_1$}$(i, j)$ $*$ \translate{$e_2$}$(i, j)$ \\
\translate{$e_1 \otimes e_2$}$(i, j)$ & $:=$ & \translate{$e_1$}$\big(i', j'\big)$ $+$ \translate{$e_2$}$\big(i'', j''\big)$ \\
\multicolumn{2}{r}{where} & $(m, n) = dims(e_2)$, \\
& &
$i'=\big\lfloor$ $i / m$ $\big\rfloor$, $j'=\big\lfloor$ $j / n$ $\big\rfloor$ \\
& &
$i''=i \% m$, $j''=j \% n$ \\
\end{tabular} &
\begin{tabular}{r c l}
\translate{$e_1 \oplus e_2$}$(i, j)$ & $:=$ & \translate{$e_1$}$(i, j)$ $+$ \translate{$e_2$} $\big(i', j' \big)$ \\
\multicolumn{2}{r}{where} & $(m, n) = dims(e_1)$\\
& &
$i'= i - m$ , $j'= j - n$ \\
\translate{$e_1 \cdot e_2$}$()$ & $:=$ & \translate{$e_1$}$(i)$ $*$ \translate{$e_2$}$(i)$ \\
\translate{$e_1 + e_2$}$(i)$ & $:=$ & \translate{$e_1$}$(i)$ $+$ \translate{$e_2$}$(i)$ \\
\translate{$e_1 \oplus e_2$}$(i)$ & $:=$ & \translate{$e_1$}$(i)$ $+$ \translate{$e_2$} $\big(i'\big)$ \\
\multicolumn{2}{r}{where} & $(n) = dims(e_1)$, $i'= i-n$\\
\translate{$e_1 \otimes e_2$}$(i, j)$ & $:=$ & \translate{$e_1$}$(i)$ $*$ \translate{$e_2$}$(j)$
\end{tabular}
\end{tabular}
\caption{Representation of linear algebra operations in \unifiedir{}. Here, $\matrixmult$ and $\otimes$  are overloaded; they correspond to matrix multiplication/Kronecker product for matrices and inner/outer product for vectors, respectively.}
\label{fig:operationIR}
\end{figure}

%% file: figures/structure-in-stur.tex
\begin{figure}

\begin{tabular}{c}
$e$: $\mathcal{Z}$
\\ \hline
$T_U(i, j) := \emptyset$ 
\end{tabular} 
\hspace{1cm}
\begin{tabular}{c}
$e$: $\mathcal{G}$
\\ \hline
$T_U(i, j) := (0 \leq i < m) * (0 \leq j < n)$ 
\end{tabular} 

\begin{tabular}{c}
$e$: $\mathcal{H}_{n1, n2}$ $\quad$ $0 \leq n1 < m$ $\quad$ $0 \leq n2 < n$
\\ \hline
$T_U(i, j) := (i = n1) * (j = n2)$ 
\end{tabular} 
\hspace{1cm}
\begin{tabular}{c}
$e$: $\mathcal{D}$ $\quad$ $m = n$
\\ \hline
$T_U(i, j) := (0 \leq i < n) * (i = j)$ 
\end{tabular} 

\begin{tabular}{c}
$e$: $\mathcal{R}_{r}$ $\quad$ $0 \leq r < m$
\\ \hline
$T_U(i, j) := (i = r) * (0 \leq j < n)$ \\
$e$: $\mathcal{C}_{c}$ $\quad$ $0 \leq c < n$
\\ \hline
$T_U(i, j) := (0 \leq i < m) * (j = c)$ 
\end{tabular} 
\hspace{1cm}
\begin{tabular}{r c l}
\multicolumn{3}{c}{$e$: $\mathcal{S}$ $\quad\quad\quad$ $m = n$}
\\ \hline
$T_U(i, j)$ &$:=$& $(0 \leq i \leq j < n)$ \\
$T_R(i, j, i', j')$ & $:=$ & $(0 \leq j < i < n)$ $*$\\
& & $(i' = j) * (j' = i)$
\end{tabular}

\caption{Representation of well-known matrix structures in \unifiedir{}. In all cases, \translate{$e$}$=T$ and thus $T_U$ and $T_R$ correspond to its unique set and redundancy map, respectively. Furthermore, we assume that $(m, n) = dims(e)$ for all cases and $T_R(i, j, i', j') := \emptyset$ unless stated explicitly.}
\label{fig:structureIR}
\end{figure}

%% file: figures/infer-unqiueset.tex
\begin{figure}
\setlength\tabcolsep{1.5pt}

\begin{tabular}{r c l}
$T(\argvec{x})$ & $:=$ & $M(\argvec{y}) * V(\argvec{z})$ $\tab$ $\argvec{x} = \argvec{y} \cup \argvec{z}$
\\ \hline
$T_U(\argvec{x})$ &$:=$& $M_U(\argvec{y}) * V_U(\argvec{z})$
\end{tabular}
\hspace{0.3cm}
\begin{tabular}{r c l}
$T(\argvec{x})$ &$:=$& $M(\argvec{x}) + V(\argvec{x})$ 
 \\ \hline
$T_U(\argvec{x})$ &$:=$& $M_U(\argvec{x}) + V_U(\argvec{x})$
\end{tabular}
\hspace{0.3cm}
\begin{tabular}{r c l}
$T(\argvec{x})$ &$:=$& $M(\argvec{y})$ $\tab$ $\argvec{x} \subseteq \argvec{y}$ 
 \\ \hline
$T_U(\argvec{x})$ &$:=$& $M_U(\argvec{y})$
\end{tabular}
\caption{Inference rules for unique set. Here we assume the redundancy maps of inputs are empty.}
\label{fig:uniqueset}
\end{figure}

%% file: figures/infer-redundancy.tex
\begin{figure}
\setlength\tabcolsep{1.5pt}
\begin{tabular}{c}
\textbf{Special cases:}
\\ \hline \hline
\vspace{0.3cm}
\begin{tabular}{r c l}
$T(\argvec{x})$ & $:=$ & $M(\argvec{y}) * M(\argvec{z})$ $\quad$ $\argvec{x} = \argvec{y} \cup \argvec{z}$ $\quad$ $\argvec{y} \cap \argvec{z} = \emptyset$ $\quad$ $\argvec{x'} = \argvec{y'} \cup \argvec{z'}$ $\quad$ $\argvec{y'} \cap \argvec{z'} = \emptyset$
\\ \hline
$T_U(\argvec{x})$ & $:=$ & $M_U(\argvec{y}) * M_U(\argvec{z}) * (\argvec{y} \leq \argvec{z})$ \\ 
$T_R(\argvec{x}, \argvec{x'})$ & $:=$ &  $M_R(\argvec{y}, \argvec{y'}) * M_R(\argvec{z}, \argvec{z'}) + 
M_U(\argvec{y}) * (\argvec{y} = \argvec{y'}) * M_R(\argvec{z}, \argvec{z'}) + $ \\
& &
$M_R(\argvec{y}, \argvec{y'}) * M_U(\argvec{z}) * (\argvec{z} = \argvec{z'}) + M_U(\argvec{y}) * (\argvec{y} = \argvec{y'}) * M_U(\argvec{z}) * (\argvec{z} = \argvec{z'}) * (\argvec{y} > \argvec{z})$
\end{tabular}
\\
\vspace{0.3cm}
\begin{tabular}{r c l}
$T(\argvec{x})$ & $:=$ & $M(\argvec{y}) * V(\argvec{z})$ $\quad$ $\argvec{x} = \argvec{y} \cup \argvec{z}$ $\quad$ $\argvec{y} \cap \argvec{z} = \emptyset$ $\quad$ $\argvec{x'} = \argvec{y'} \cup \argvec{z'}$ $\quad$ $\argvec{y'} \cap \argvec{z'} = \emptyset$
\\ \hline
$T_U(\argvec{x})$ &$:=$& $M_U(\argvec{y}) * V_U(\argvec{z})$ \\ 
$T_R(\argvec{x}, \argvec{x'})$ &$:=$& $M_R(\argvec{y}, \argvec{y'}) * V_R(\argvec{z}, \argvec{z'}) + 
M_U(\argvec{y}) * (\argvec{y} = \argvec{y'}) * V_R(\argvec{z}, \argvec{z'}) + M_R(\argvec{y}, \argvec{y'}) * V_U(\argvec{z}) *  (\argvec{z} = \argvec{z'})$ 
\end{tabular}
\\

\\
\vspace{0.3cm}

\begin{tabular}{r c l}
$T(\argvec{x})$ &$:=$& $M(\argvec{x}) + V(\argvec{x})$ $\quad$ $M_U(\argvec{x}) = V_U(\argvec{x})$ $\quad$ $M_R(\argvec{x}, \argvec{x'}) = V_R(\argvec{x}, \argvec{x'})$
 \\ \hline
$T_U(\argvec{x})$ &$:=$& $M_U(\argvec{x})$ \\ 
$T_R(\argvec{x}, \argvec{x'})$ &$:=$& $M_R(\argvec{x}, \argvec{x'})$
\end{tabular}

\\
\vspace{0.3cm}

\begin{tabular}{r c l}
$T(\argvec{x})$ &$:=$& $M(\argvec{x}) + V(\argvec{y})*(\argvec{y}=\argvec{x}-\argvec{d})$ $\quad$ $\argvec{d} = dim(M)$
 \\ \hline
$T_U(\argvec{x})$ &$:=$& $M_U(\argvec{x}) + V_U(\argvec{y})*(\argvec{y}=\argvec{x}-\argvec{d})$ \\ 
$T_R(\argvec{x}, \argvec{x'})$ &$:=$& $M_R(\argvec{x}, \argvec{x'}) + 
V_R(\argvec{y}, \argvec{y'})*(\argvec{y}=\argvec{x}-\argvec{d})*(\argvec{y'}=\argvec{x'}-\argvec{d})$ 
\end{tabular}

\\
\vspace{0.3cm}

\begin{tabular}{r c l}
$T(\argvec{x})$ & $:=$ & $\argvec{b} \leq \argvec{x} < \argvec{c}$ 
\\ \hline
$T_U(\argvec{x})$ &$:=$& $\argvec{x} = \argvec{b}$ \\ 
$T_R(\argvec{x}, \argvec{x'})$ &$:=$& $((\argvec{b} \leq \argvec{x} < \argvec{c}) - (\argvec{b} = \argvec{x})) * (\argvec{x'} = \argvec{b})$
\end{tabular}

\\
\textbf{General cases}:
\\ \hline \hline
\vspace{0.3cm}
\begin{tabular}{r c l}
$T(\argvec{x})$ & $:=$ & $M(\argvec{y}) * V(\argvec{z})$ $\quad$ $\argvec{x} = \argvec{y} \cup \argvec{z}$ $\quad$ $\argvec{x'} = \argvec{y'} \cup \argvec{z'}$
\\ \hline
$T_U(\argvec{x})$ &$:=$& $M_U(\argvec{y}) * V_U(\argvec{z}) + 
M_U(\argvec{y}) * V_R(\argvec{z}, \argvec{z'}) + M_R(\argvec{y}, \argvec{y'}) * V_U(\argvec{z})$ \\ 
$T_R(\argvec{x}, \argvec{x'})$ &$:=$& $M_R(\argvec{y}, \argvec{y'}) * V_R(\argvec{z}, \argvec{z'})$ 
\end{tabular}

\\
\vspace{0.3cm}

\begin{tabular}{r c l}
$T(\argvec{x})$ &$:=$& $M(\argvec{x}) + V(\argvec{x})$ 
 \\ \hline
$T_U(\argvec{x})$ &$:=$& $M_U(\argvec{x}) + M_R(\argvec{x}, \argvec{x'}) + V_U(\argvec{x}) + V_R(\argvec{x}, \argvec{x'})$ \\ 
$T_R(\argvec{x}, \argvec{x'})$ &$:=$& $\emptyset$ 
\end{tabular}
\hspace{0.3cm}
\begin{tabular}{r c l}
$T(\argvec{x})$ &$:=$& $M(\argvec{y})$ $\quad$ $\argvec{x} \subseteq \argvec{y}$ 
 \\ \hline
$T_U(\argvec{x})$ &$:=$& $M_U(\argvec{y}) + M_R(\argvec{y}, \argvec{y'})$ \\ 
$T_R(\argvec{x}, \argvec{x'})$ &$:=$& $\emptyset$ 
\end{tabular}

\end{tabular}

\caption{Inference rules for unique sets and redundancy maps. The priority for the rules are top down. The default case, considers $T_R(\argvec{x}, \argvec{x'}) := \emptyset$ and $T_U(\argvec{x}) := \forget{T(x)}$ (cf. Section~\ref{sec:soundness}). The last three rules subsume the ones shown in Figure~\ref{fig:uniqueset}. 
}
\label{fig:redundancymap}
\end{figure}

%% file: semantics.tex
\section{Soundness of structure inference}
\label{sec:soundness}

We present a simple and more mathematical view of what unique sets and redundancy maps compute. We then derive several properties that they satisfy. Furthermore, we state a soundness theorem for our structure inference, formally anchoring the fact that they do not lose information and compute the same as the original tensor.

\subsection{Abstract view on the reduction problem}

Abstractly, we can view the problem as follows.
Let $\TT_{n_1,\ldots,n_k}(\RR)$ be the real vector-space of $n_1\times \ldots\times n_k$ real tensors.
Given a fixed $T\in \TT_{n_1,\ldots,n_k}(\RR)$,
we study the problem of finding a pair of linear maps $(R_T, P_T):\TT_{n_1,\ldots,n_k}(\RR)\to \TT_{n_1,\ldots,n_k}(\RR)$ such that the following simple equation holds
\begin{equation}
\label{eqn:problem-statement}
    R_T\circ P_T(T)=T
\end{equation} 
We call such a pair $(R_T, P_T)$ a solution to the reduction problem for $T$.
The intuition is that $P_T$ represents a projection, $P_T(T)$ represents a compressed version of $T$, and $R_T$ ensures that this compression is lossless. 
Note that there are a lot of solutions to this problem, e.g. $P_T$ could simply be the identity, and we are naturally interested in solutions that approximately minimize the number of non-zeros elements of the compressed tensor $P_T(T)$. Finding optimal solutions is also feasible but would make us lose time overall for the final computations we are interested in.

Note that, as $P_T, R_T$ are linear, they can be represented by $n_1\times \ldots\times n_k \times n_1 \times\ldots \times n_k$ tensors.  In this work, we will restrict to the cases where $P_T, R_T$ can be represented with tensors valued in $\{0,1\}$ and both satisfy that, when seen as matrices, the sum of the elements on each row is at most 1. Such matrices are called partial injective relations.  
We further ask $P_T$ to be a projection.
In such a case, $P_T$ is an orthogonal projection and therefore verifies $P_T\circ P_T=P_T$ (an orthogonal projection is automatically a partial injective relation).
Any partial injective relation $F$ satisfies the nice property that $F(A\odot B)=F(A)\odot F(B)$ for all $A, B$, where $\odot$ is the Hadamard product. In fact, this is a characterization of partial injective relations.

We denote by $\forget{-}:\TT_{n_1,\ldots,n_k}(\RR)\to\TT_{n_1,\ldots,n_k}(\RR)$ the function sending a real $r$ at position $(i_1,\ldots,i_k)$ to $1$ at the same position if $r\neq 0$ and to $0$ otherwise. Note that this extends a monoid homomorphism $(\RR,\times,1)\to (\{0,1\},\times,1)$ but this is not a linear transformation as it does not commute with addition. 
It still implies that $\forget{T\otimes S}=\forget{T}\otimes \forget{S}$,  $\forget{T\oplus S}=\forget{T}\oplus \forget{S}$ and $\forget{T\odot S}=\forget{T}\odot \forget{S}$, where $\otimes$ is the Kronecker product, and $\oplus$ the direct sum. Additionally, for any partial injective relation $F$ and any $A$, we have $\forget{F(A)}=F(\forget{A})$. 

We indistinguishably use tensors and the linear maps they represent, assuming fixed bases of the underlying vector spaces at play.
From the abstract setting, the tensors $T_U, T_R, T_C$ we infer and use in our unified IR are derived as follows:
\begin{equation}
\label{eqn:tensors-from-abstract}
     T_U:= \forget{P_T(T)} \quad\quad\quad
    T_R:= R_T-P_T\quad\quad\quad
    T_C := P_T(T) 
\end{equation}
In other words, $T_U$ is the support of the compressed tensor $T_C = P_T(T)$, and $T_R$ is, up to a small optimization, the support of the tensor representation of $R_T$. From Equation~\ref{eqn:problem-statement} and the definitions of $T_U, T_R, T_C$ (Equation~\ref{eqn:tensors-from-abstract}), we obtain several properties that are key in reconstructing $T$ and for optimisations. 

\smartpara{Proposition} The following Properties are valid equations, where $\argvec{x}$ is the list of free variables in each case:
\begin{enumerate}
    \item $T_U(\argvec{x})* T_R(\argvec{x}, \argvec{x'}) = \emptyset$. This optimisation ensures that unique elements should not be mapped to any other element by $T_R$.
    \item $T_C(\argvec{x}) = T_U(\argvec{x}) * T(\argvec{x})$ is the equation defining the compressed tensor in our IR.
    \item $ T_C(\argvec{x}) * T_R(\argvec{x}, \argvec{x'})= \emptyset$. Elements that are stored in the compressed format should only exist in the unique set.
    \item $T_U(\argvec{x}) * T_U(\argvec{x}) = T_U(\argvec{x})$. This is another simple optimisation.
    \item $T(\argvec{x}) = T_C(\argvec{x}) + T_R(\argvec{x}, \argvec{x'}) * T_C(\argvec{x'})$. This is the reconstruction process of the tensor $T$ given its compressed format $T_C$, its unique set $T_U$, and its redundancy map $T_R$. This captures the key lossless property of the contraction. 
    \label{property-last}
\end{enumerate}

\smartpara{Proof sketch} 
Note the following extra elementary properties that we will use in the proofs.
$A\odot \forget{A} = A$ for any $A$ and $P(A)\odot B= A\odot P(B) = P(A)\odot P(B)$ for any $A, B$ and orthogonal projection $P$ (which is a restatement of a well-known characterization of orthogonal projections in terms of inner products).

(2) \vspace{-.6cm} 
\begin{align*}
    T\odot T_U &= T\odot \forget{P_T(T)} &\text{definition of $T_U$} \\
 &= T\odot P_T(\forget{T}) &\text{property of partial injective relation $P_T$} \\
 &= P_T(T) \odot P_T(\forget{T}) &\text{Property of orthogonal projection $P_T$} \\
 &= P_T(T\odot \forget{T}) &\text{property of partial injective relation $P_T$} \\
 &= P_T(T) = T_C &\text{fundamental property of $\forget{-}$, definition of $T_C$}
\end{align*}

(3) is easily obtained from (1) and the fact that $T_R$ commutes with $\forget{-}$.

(4)  \vspace{-.6cm} 
\begin{align*}
    T_U\odot T_U &= \forget{P_T(T)}\odot \forget{P_T(T)}  &\text{definition of $T_U$}\\
    &=P_T(\forget{T})\odot P_T(\forget{T})  &\text{property of partial injective relation $P_T$} \\
    &= P_T(\forget{T} \odot \forget{T})  &\text{property of partial injective relation $P_T$} \\
    &= P_T(\forget{T})= T_U  &\text{property of $\forget{-}$, definition of $T_U$} 
\end{align*}

(5)  \vspace{-.6cm} 
\begin{align*}
    T_C + T_R(T_C) &= P_T(T) + (R_T-P_T)(P_T(T)) &\text{definition of $T_R,T_C$}\\
    &= P_T(T) + R_TP_T(T) - P_TP_T(T) &\text{linearity} \\
    &= P_T(T) + T - P_T(T) = T &\text{Equation~\ref{eqn:problem-statement}, $P_TP_T=P_T$, simplification}
\end{align*} \qed

\subsection{Soundness results}

Given a tensor $T$, we write $\II_T:(R_T,P_T)\to (T_U,T_R,T_C)$ for the mapping sending a pair of linear maps verifying Equation~\ref{eqn:problem-statement} to the tensors defined by Equation~\ref{eqn:tensors-from-abstract}. We say that $(T_U,T_R,T_C)$ implements the solution $(R_T,P_T)$ to the reduction problem for $T$.

In this work, we never explicitly construct $P_T$ and $R_T$, but we do inductively define implementations $(T_U, T_R, T_C)$ of solutions to the reduction problem for $T$, by induction on the structure of $T$. 
The first natural question is therefore whether the inferred tuple $(T_U, T_R, T_C)$ is sound, that is whether is obtained as an implementation of a solution for the reduction problem for $T$. This is a sufficient condition, as by Property~\ref{property-last} above we can reconstruct $T$ from such a tuple $(T_U, T_C, T_R)$.

\smartpara{Theorem}
Let $M,N$ be tensors. Assume given pairs $(R_M,P_M)$ and $(R_N,P_N)$ that are solutions to the reduction problem for $M$ and $N$, respectively. Let $(M_U, M_R, M_C)$ and $(N_U, N_R, N_C)$ be their respective implementations. 
Further, assume that $T$ is given by the premise of an inference rule from Figure~\ref{fig:redundancymap}. 
Then, there exists a solution $(R_T, P_T)$ for the reduction problem for $T$ such that its implementation $(T_U, T_R, T_C)$ is given by the conclusion of the inference rule corresponding to $T$'s definition.

\smartpara{Proof sketch}
We sketch the proof for some of the important cases.
For the second rule, let $P_T:=P_M\otimes P_V$ and $R_T:=R_M\otimes R_T$.
Then $R_TP_T(T)= (R_M\otimes R_T)(P_M\otimes P_T)(M\otimes V)= R_MP_M(M)\otimes R_VP_V(V)=M\otimes V=T$. In addition, $\forget{P_T(T)}=\forget{P_M(M)\otimes P_V(V)}=\forget{P_M(M)}\otimes \forget{P_V(V)}=M_U\otimes V_U=T_U$.
Finally, 
\begin{align*}
    R_T-P_T &= R_M\otimes R_V-P_M\otimes P_V \\
    & =(M_R+P_M)\otimes (V_R+P_V)-P_M\otimes P_V \\
    &= M_R\otimes V_R + P_M\otimes V_R + M_R\otimes P_V
\end{align*}
Now, note that we define $P_M$ to be the orthogonal projections onto $M_U$, i.e. $P_M$ can be represented as $P_M = M_U\odot (-)$. This can be represented in \unifiedir{} as $M_U(\argvec{z})*(z=z')$. The first rule is proved from the second by noting that $(M\otimes M)_{im+i',jn+j'}=M_{i,j}*M_{i',j'}=M_{i',j'}*M_{i,j}=(M\otimes M)_{i'm+i,j'n+j}$ when $M$ is an $m\times n$ matrix. This immediately generalizes to the case where $M$ is an arbitrary tensor. The third rule is straightforward. 
For the fourth rule, we have $T=M\oplus V$. Let $R_T:=R_M\oplus R_V$ and $P_T:=P_M\oplus P_V$. The remainder of the proof is the same as for rule 2, where we replace $\otimes$ by $\oplus$.

\qed

As a direct corollary of the theorem above, we obtain the soundness of our structure inference.

\smartpara{Corollary [Soundness of Inference]}
Let $T:=f(M, V)$ be given by a premise of an inference rule from Figure~\ref{fig:redundancymap}. Assume Property~\ref{property-last} holds for $M, V$. Then, Property~\ref{property-last} holds for $T$, where $T_U, T_R$ are computed by the conclusion of the same inference rule.

A key rewrite we use in the optimization phase is inlining definitions. This operation is sound in our language.

\smartpara{Proposition [Substitution Lemma]}
Let $T_1(\argvec{x}):= B_1(\argvec{x,x'})$ and $T_2 (\argvec{y}):= B_2(\argvec{y,y'})$.\\
Then $T_2(\argvec{y}) := B_2[B_1/T_1](\argvec{x, x',y'})$ is semantically valid whenever $\argvec{x'}\cap \argvec{y'}=\argvec{x}\cap\argvec{y'}=\emptyset$.

%% file: related.tex
\section{Related Work}
\label{sec:related}

\smartpara{Dense Tensor Algebra}
Polyhedral frameworks such as isl~\cite{VERDOOLAEGE2010ISL} and CLooG~\cite{bastoul2004code} provide efficient loop nest code generation for dense and affine tensor algebra computations. Tensor Comprehension~\cite{vasilache2018tensor} provides a polyhedral-based DSL supporting generalized Einstein notation that leads to optimized Cuda code generation for dense deep learning computation. TensorFlow~\cite{tensorflow2015-whitepaper} provides dense tensor operation for large-scale machine learning applications.\footnote{There is limited support for sparse processing in TensorFlow~\cite{tensorflow2015-whitepaper}, but it was shown to be suboptimal~\cite{chou2018format}.} 
All these works provide efficient kernels for dense linear and tensor algebra but for structured tensors, still, they do unnecessary and/or redundant computations.

\smartpara{Sparse Tensor Algebra} The sparse polyhedral framework~\cite{strout2018sparse} extends the ability of polyhedral compilation to support sparse tensor algebra as well. TACO~\cite{kjolstad:2017:taco} handles sparse and dense computation over tensor algebra. However, unlike \systemname{}, none of these work support redundancy-aware computation. Moreover, sparsity is handled in run-time that leads to irregular memory access that are hard to optimize for the compiler~\cite{DBLP:journals/cgf/TangSKPLP20}.

\smartpara{Specialized Sparse Linear Algebra}
\cite{augustine2019generating} take a different approach by breaking the irregular sparsity patterns into sub-computations with regular structures so they can remove indirect access and provide vectorization to the linear algebra code. 
EGGS~\cite{DBLP:journals/cgf/TangSKPLP20} further specializes the computation to a sparsity pattern and creates the expression tree of the result by unrolling the entire computing. Performing common-subexpression over the expression tree can partially remove redundancies, but cannot detect symmetric-style patterns. 
\systemname{} infers the redundancy patterns at compilation time to specialize the generated code.

\smartpara{Structured Linear Algebra}
LGen~\cite{spampinato2016basic} proposes a polyhedral-based technique for code generation of small-scale structured linear algebra. 
The unique elements are provided in \texttt{SInfo}, and the
redundancy information is kept in \texttt{AInfo}. However, it does not support higher-order tensor computations and is limited to fixed-size small-scale matrices.
To the best of our knowledge, there is no previous work supporting the structure for higher-order tensors.

\smartpara{Declarative Data Processing Languages} \unifiedir{} is closely connected to logic languages such as Datalog and Prolog.
There are two main differences. First, \unifiedir{} does not allow recursive definitions. 
Second, in addition to sets, \unifiedir{} allows for aggregations over tensors. 
Dyna~\cite{DBLP:conf/datalog/EisnerF10} extends Datalog by adding to support for maps to real numbers rather than only supporting sets, which are maps to boolean values.
FAQ~\cite{DBLP:conf/pods/KhamisNR16} is the main source of inspiration for \unifiedir{}, which allows for a combination of different semi-rings.
The main focus for FAQ has been on efficient algorithms for evaluating sparse tensor contractions appearing in database query engines~\cite{khamis2020learning} (e.g., worst-case optimal joins).
\unifiedir{} provides additional constructs for arithmetic operations over the indices and restricting the ranges, which are crucial for efficient structured tensor computations.

\begin{table}
    \centering
    \begin{tabular}{|c|c|c|c|c|c|c|}
        \hline
        \textbf{Framework} & \textbf{LA} & \textbf{TA} & \textbf{Dense} & \textbf{Sparse} & \textbf{Redundancy} & \textbf{Loop Opts.} \\ \hline
        Dense TA (TensorFlow) & \supfull{} & \supfull{} & \supfull{} & \supnone{} & \supnone{} & \supfull{} \\ \hline
        Dense LA (BLAS) & \supfull{} & \supnone{} & \supfull{} & \supnone{} & \supnone{} & \supfull{} \\ \hline
        Sparse TA (TACO, SPLATT) & \supfull{} & \supfull{} & \supfull{} & \supfull{}  & \supnone{} & \supfull{} \\ \hline
        Sparse LA (MKL, OSKI) & \supfull{} & \supnone{} & \supfull{} & \supfull{}  & \supnone{} & \supfull{} \\ \hline
        Static Sparse LA (EGGS) & \supfull{} & \supnone{} & \supfull{} & \suphalf{} (FS) & \suphalf{} &  \supfull{} \\ \hline
        Structured LA (LGen) & \supfull{} & \supnone{} & \supfull{} & \suphalf{} (FS) & \supfull{} & \supfull{} \\ \hline
        \systemname{} & \supfull{} & \supfull{} & \supfull{} & \suphalf{} (SY) & \supfull{} & \suphalf \\ \hline
    \end{tabular}
    \caption{Tensor algebra frameworks comparison. FS: Fixed-size, SY: Symbolic}
    \label{tbl:related}
\end{table}

%% file: main.bbl

\begin{thebibliography}{31}


\ifx \showCODEN    \undefined \def \showCODEN     #1{\unskip}     \fi
\ifx \showDOI      \undefined \def \showDOI       #1{#1}\fi
\ifx \showISBNx    \undefined \def \showISBNx     #1{\unskip}     \fi
\ifx \showISBNxiii \undefined \def \showISBNxiii  #1{\unskip}     \fi
\ifx \showISSN     \undefined \def \showISSN      #1{\unskip}     \fi
\ifx \showLCCN     \undefined \def \showLCCN      #1{\unskip}     \fi
\ifx \shownote     \undefined \def \shownote      #1{#1}          \fi
\ifx \showarticletitle \undefined \def \showarticletitle #1{#1}   \fi
\ifx \showURL      \undefined \def \showURL       {\relax}        \fi
\providecommand\bibfield[2]{#2}
\providecommand\bibinfo[2]{#2}
\providecommand\natexlab[1]{#1}
\providecommand\showeprint[2][]{arXiv:#2}

\bibitem[Abadi et~al\mbox{.}(2015)]%
        {tensorflow2015-whitepaper}
\bibfield{author}{\bibinfo{person}{Mart\'{i}n Abadi}, \bibinfo{person}{Ashish
  Agarwal}, \bibinfo{person}{Paul Barham}, \bibinfo{person}{Eugene Brevdo},
  \bibinfo{person}{Zhifeng Chen}, \bibinfo{person}{Craig Citro},
  \bibinfo{person}{Greg~S. Corrado}, \bibinfo{person}{Andy Davis},
  \bibinfo{person}{Jeffrey Dean}, \bibinfo{person}{Matthieu Devin},
  \bibinfo{person}{Sanjay Ghemawat}, \bibinfo{person}{Ian Goodfellow},
  \bibinfo{person}{Andrew Harp}, \bibinfo{person}{Geoffrey Irving},
  \bibinfo{person}{Michael Isard}, \bibinfo{person}{Yangqing Jia},
  \bibinfo{person}{Rafal Jozefowicz}, \bibinfo{person}{Lukasz Kaiser},
  \bibinfo{person}{Manjunath Kudlur}, \bibinfo{person}{Josh Levenberg},
  \bibinfo{person}{Dandelion Man\'{e}}, \bibinfo{person}{Rajat Monga},
  \bibinfo{person}{Sherry Moore}, \bibinfo{person}{Derek Murray},
  \bibinfo{person}{Chris Olah}, \bibinfo{person}{Mike Schuster},
  \bibinfo{person}{Jonathon Shlens}, \bibinfo{person}{Benoit Steiner},
  \bibinfo{person}{Ilya Sutskever}, \bibinfo{person}{Kunal Talwar},
  \bibinfo{person}{Paul Tucker}, \bibinfo{person}{Vincent Vanhoucke},
  \bibinfo{person}{Vijay Vasudevan}, \bibinfo{person}{Fernanda Vi\'{e}gas},
  \bibinfo{person}{Oriol Vinyals}, \bibinfo{person}{Pete Warden},
  \bibinfo{person}{Martin Wattenberg}, \bibinfo{person}{Martin Wicke},
  \bibinfo{person}{Yuan Yu}, {and} \bibinfo{person}{Xiaoqiang Zheng}.}
  \bibinfo{year}{2015}\natexlab{}.
\newblock \bibinfo{title}{{TensorFlow}: Large-Scale Machine Learning on
  Heterogeneous Systems}.
\newblock
\newblock
\urldef\tempurl%
\url{https://www.tensorflow.org/}
\showURL{%
\tempurl}
\newblock
\shownote{Software available from tensorflow.org}.


\bibitem[Augustine et~al\mbox{.}(2019)]%
        {augustine2019generating}
\bibfield{author}{\bibinfo{person}{Travis Augustine},
  \bibinfo{person}{Janarthanan Sarma}, \bibinfo{person}{Louis{-}No{\"{e}}l
  Pouchet}, {and} \bibinfo{person}{Gabriel Rodr{\'{\i}}guez}.}
  \bibinfo{year}{2019}\natexlab{}.
\newblock \showarticletitle{Generating piecewise-regular code from irregular
  structures}. In \bibinfo{booktitle}{\emph{Proceedings of the 40th {ACM}
  {SIGPLAN} Conference on Programming Language Design and Implementation,
  {PLDI} 2019, Phoenix, AZ, USA, June 22-26, 2019}},
  \bibfield{editor}{\bibinfo{person}{Kathryn~S. McKinley} {and}
  \bibinfo{person}{Kathleen Fisher}} (Eds.). \bibinfo{publisher}{{ACM}},
  \bibinfo{pages}{625--639}.
\newblock
\urldef\tempurl%
\url{https://doi.org/10.1145/3314221.3314615}
\showDOI{\tempurl}


\bibitem[Bastoul(2004)]%
        {bastoul2004code}
\bibfield{author}{\bibinfo{person}{C{\'{e}}dric Bastoul}.}
  \bibinfo{year}{2004}\natexlab{}.
\newblock \showarticletitle{Code Generation in the Polyhedral Model Is Easier
  Than You Think}. In \bibinfo{booktitle}{\emph{13th International Conference
  on Parallel Architectures and Compilation Techniques {(PACT} 2004), 29
  September - 3 October 2004, Antibes Juan-les-Pins, France}}.
  \bibinfo{publisher}{{IEEE} Computer Society}, \bibinfo{pages}{7--16}.
\newblock
\urldef\tempurl%
\url{https://doi.org/10.1109/PACT.2004.10018}
\showDOI{\tempurl}


\bibitem[Chou et~al\mbox{.}(2018)]%
        {chou2018format}
\bibfield{author}{\bibinfo{person}{Stephen Chou}, \bibinfo{person}{Fredrik
  Kjolstad}, {and} \bibinfo{person}{Saman Amarasinghe}.}
  \bibinfo{year}{2018}\natexlab{}.
\newblock \showarticletitle{Format abstraction for sparse tensor algebra
  compilers}.
\newblock \bibinfo{journal}{\emph{Proceedings of the ACM on Programming
  Languages}} \bibinfo{volume}{2}, \bibinfo{number}{OOPSLA}
  (\bibinfo{year}{2018}), \bibinfo{pages}{1--30}.
\newblock


\bibitem[Cichocki et~al\mbox{.}(2009)]%
        {cichocki2009nonnegative}
\bibfield{author}{\bibinfo{person}{Andrzej Cichocki}, \bibinfo{person}{Rafal
  Zdunek}, \bibinfo{person}{Anh~Huy Phan}, {and} \bibinfo{person}{Shun-ichi
  Amari}.} \bibinfo{year}{2009}\natexlab{}.
\newblock \bibinfo{booktitle}{\emph{Nonnegative matrix and tensor
  factorizations: applications to exploratory multi-way data analysis and blind
  source separation}}.
\newblock \bibinfo{publisher}{John Wiley \& Sons}.
\newblock


\bibitem[Dongarra et~al\mbox{.}(1990)]%
        {DBLP:journals/toms/DongarraCHD90}
\bibfield{author}{\bibinfo{person}{Jack~J. Dongarra},
  \bibinfo{person}{Jeremy~Du Croz}, \bibinfo{person}{Sven Hammarling}, {and}
  \bibinfo{person}{Iain~S. Duff}.} \bibinfo{year}{1990}\natexlab{}.
\newblock \showarticletitle{A set of level 3 basic linear algebra subprograms}.
\newblock \bibinfo{journal}{\emph{{ACM} Trans. Math. Softw.}}
  \bibinfo{volume}{16}, \bibinfo{number}{1} (\bibinfo{year}{1990}),
  \bibinfo{pages}{1--17}.
\newblock
\urldef\tempurl%
\url{https://doi.org/10.1145/77626.79170}
\showDOI{\tempurl}


\bibitem[Eisner and Filardo(2010)]%
        {DBLP:conf/datalog/EisnerF10}
\bibfield{author}{\bibinfo{person}{Jason Eisner} {and}
  \bibinfo{person}{Nathaniel~Wesley Filardo}.} \bibinfo{year}{2010}\natexlab{}.
\newblock \showarticletitle{Dyna: Extending Datalog for Modern {AI}}. In
  \bibinfo{booktitle}{\emph{Datalog Reloaded - First International Workshop,
  Datalog 2010, Oxford, UK, March 16-19, 2010. Revised Selected Papers}}
  \emph{(\bibinfo{series}{Lecture Notes in Computer Science},
  Vol.~\bibinfo{volume}{6702})}, \bibfield{editor}{\bibinfo{person}{Oege
  de~Moor}, \bibinfo{person}{Georg Gottlob}, \bibinfo{person}{Tim Furche},
  {and} \bibinfo{person}{Andrew~Jon Sellers}} (Eds.).
  \bibinfo{publisher}{Springer}, \bibinfo{pages}{181--220}.
\newblock
\urldef\tempurl%
\url{https://doi.org/10.1007/978-3-642-24206-9\_11}
\showDOI{\tempurl}


\bibitem[Favorita(2017)]%
        {favorita}
\bibfield{author}{\bibinfo{person}{Corporacion Favorita}.}
  \bibinfo{year}{2017}\natexlab{}.
\newblock \bibinfo{title}{{Corp. Favorita Grocery Sales Forecasting: Can you
  accurately predict sales for a large grocery chain?}}
\newblock
\newblock


\bibitem[Gareev et~al\mbox{.}(2018)]%
        {DBLP:journals/taco/GareevGK18}
\bibfield{author}{\bibinfo{person}{Roman Gareev}, \bibinfo{person}{Tobias
  Grosser}, {and} \bibinfo{person}{Michael Kruse}.}
  \bibinfo{year}{2018}\natexlab{}.
\newblock \showarticletitle{High-Performance Generalized Tensor Operations: {A}
  Compiler-Oriented Approach}.
\newblock \bibinfo{journal}{\emph{{ACM} Trans. Archit. Code Optim.}}
  \bibinfo{volume}{15}, \bibinfo{number}{3} (\bibinfo{year}{2018}),
  \bibinfo{pages}{34:1--34:27}.
\newblock
\urldef\tempurl%
\url{https://doi.org/10.1145/3235029}
\showDOI{\tempurl}


\bibitem[Harris et~al\mbox{.}(2020)]%
        {harris2020array}
\bibfield{author}{\bibinfo{person}{Charles~R. Harris},
  \bibinfo{person}{K.~Jarrod Millman}, \bibinfo{person}{St{\'{e}}fan~J. van~der
  Walt}, \bibinfo{person}{Ralf Gommers}, \bibinfo{person}{Pauli Virtanen},
  \bibinfo{person}{David Cournapeau}, \bibinfo{person}{Eric Wieser},
  \bibinfo{person}{Julian Taylor}, \bibinfo{person}{Sebastian Berg},
  \bibinfo{person}{Nathaniel~J. Smith}, \bibinfo{person}{Robert Kern},
  \bibinfo{person}{Matti Picus}, \bibinfo{person}{Stephan Hoyer},
  \bibinfo{person}{Marten~H. van Kerkwijk}, \bibinfo{person}{Matthew Brett},
  \bibinfo{person}{Allan Haldane}, \bibinfo{person}{Jaime~Fern{\'{a}}ndez del
  R{\'{i}}o}, \bibinfo{person}{Mark Wiebe}, \bibinfo{person}{Pearu Peterson},
  \bibinfo{person}{Pierre G{\'{e}}rard-Marchant}, \bibinfo{person}{Kevin
  Sheppard}, \bibinfo{person}{Tyler Reddy}, \bibinfo{person}{Warren Weckesser},
  \bibinfo{person}{Hameer Abbasi}, \bibinfo{person}{Christoph Gohlke}, {and}
  \bibinfo{person}{Travis~E. Oliphant}.} \bibinfo{year}{2020}\natexlab{}.
\newblock \showarticletitle{Array programming with {NumPy}}.
\newblock \bibinfo{journal}{\emph{Nature}} \bibinfo{volume}{585},
  \bibinfo{number}{7825} (\bibinfo{date}{Sept.} \bibinfo{year}{2020}),
  \bibinfo{pages}{357--362}.
\newblock
\urldef\tempurl%
\url{https://doi.org/10.1038/s41586-020-2649-2}
\showDOI{\tempurl}


\bibitem[Hegde et~al\mbox{.}(2019)]%
        {DBLP:conf/micro/HegdeMPCJSEF19}
\bibfield{author}{\bibinfo{person}{Kartik Hegde}, \bibinfo{person}{Hadi~Asghari
  Moghaddam}, \bibinfo{person}{Michael Pellauer}, \bibinfo{person}{Neal~Clayton
  Crago}, \bibinfo{person}{Aamer Jaleel}, \bibinfo{person}{Edgar Solomonik},
  \bibinfo{person}{Joel~S. Emer}, {and} \bibinfo{person}{Christopher~W.
  Fletcher}.} \bibinfo{year}{2019}\natexlab{}.
\newblock \showarticletitle{ExTensor: An Accelerator for Sparse Tensor
  Algebra}. In \bibinfo{booktitle}{\emph{Proceedings of the 52nd Annual
  {IEEE/ACM} International Symposium on Microarchitecture, {MICRO} 2019,
  Columbus, OH, USA, October 12-16, 2019}}. \bibinfo{publisher}{{ACM}},
  \bibinfo{pages}{319--333}.
\newblock
\urldef\tempurl%
\url{https://doi.org/10.1145/3352460.3358275}
\showDOI{\tempurl}


\bibitem[Hirata(2003)]%
        {doi:10.1021/jp034596z}
\bibfield{author}{\bibinfo{person}{So Hirata}.}
  \bibinfo{year}{2003}\natexlab{}.
\newblock \showarticletitle{Tensor Contraction Engine: Abstraction and
  Automated Parallel Implementation of Configuration-Interaction,
  Coupled-Cluster, and Many-Body Perturbation Theories}.
\newblock \bibinfo{journal}{\emph{The Journal of Physical Chemistry A}}
  \bibinfo{volume}{107}, \bibinfo{number}{46} (\bibinfo{year}{2003}),
  \bibinfo{pages}{9887--9897}.
\newblock
\urldef\tempurl%
\url{https://doi.org/10.1021/jp034596z}
\showDOI{\tempurl}


\bibitem[Hirata(2006)]%
        {hirata2006symbolic}
\bibfield{author}{\bibinfo{person}{So Hirata}.}
  \bibinfo{year}{2006}\natexlab{}.
\newblock \showarticletitle{Symbolic algebra in quantum chemistry}.
\newblock \bibinfo{journal}{\emph{Theoretical Chemistry Accounts}}
  \bibinfo{volume}{116}, \bibinfo{number}{1} (\bibinfo{year}{2006}),
  \bibinfo{pages}{2--17}.
\newblock


\bibitem[Jouppi et~al\mbox{.}(2017)]%
        {DBLP:conf/isca/JouppiYPPABBBBB17}
\bibfield{author}{\bibinfo{person}{Norman~P. Jouppi}, \bibinfo{person}{Cliff
  Young}, \bibinfo{person}{Nishant Patil}, \bibinfo{person}{David~A.
  Patterson}, \bibinfo{person}{Gaurav Agrawal}, \bibinfo{person}{Raminder
  Bajwa}, \bibinfo{person}{Sarah Bates}, \bibinfo{person}{Suresh Bhatia},
  \bibinfo{person}{Nan Boden}, \bibinfo{person}{Al Borchers},
  \bibinfo{person}{Rick Boyle}, \bibinfo{person}{Pierre{-}luc Cantin},
  \bibinfo{person}{Clifford Chao}, \bibinfo{person}{Chris Clark},
  \bibinfo{person}{Jeremy Coriell}, \bibinfo{person}{Mike Daley},
  \bibinfo{person}{Matt Dau}, \bibinfo{person}{Jeffrey Dean},
  \bibinfo{person}{Ben Gelb}, \bibinfo{person}{Tara~Vazir Ghaemmaghami},
  \bibinfo{person}{Rajendra Gottipati}, \bibinfo{person}{William Gulland},
  \bibinfo{person}{Robert Hagmann}, \bibinfo{person}{C.~Richard Ho},
  \bibinfo{person}{Doug Hogberg}, \bibinfo{person}{John Hu},
  \bibinfo{person}{Robert Hundt}, \bibinfo{person}{Dan Hurt},
  \bibinfo{person}{Julian Ibarz}, \bibinfo{person}{Aaron Jaffey},
  \bibinfo{person}{Alek Jaworski}, \bibinfo{person}{Alexander Kaplan},
  \bibinfo{person}{Harshit Khaitan}, \bibinfo{person}{Daniel Killebrew},
  \bibinfo{person}{Andy Koch}, \bibinfo{person}{Naveen Kumar},
  \bibinfo{person}{Steve Lacy}, \bibinfo{person}{James Laudon},
  \bibinfo{person}{James Law}, \bibinfo{person}{Diemthu Le},
  \bibinfo{person}{Chris Leary}, \bibinfo{person}{Zhuyuan Liu},
  \bibinfo{person}{Kyle Lucke}, \bibinfo{person}{Alan Lundin},
  \bibinfo{person}{Gordon MacKean}, \bibinfo{person}{Adriana Maggiore},
  \bibinfo{person}{Maire Mahony}, \bibinfo{person}{Kieran Miller},
  \bibinfo{person}{Rahul Nagarajan}, \bibinfo{person}{Ravi Narayanaswami},
  \bibinfo{person}{Ray Ni}, \bibinfo{person}{Kathy Nix},
  \bibinfo{person}{Thomas Norrie}, \bibinfo{person}{Mark Omernick},
  \bibinfo{person}{Narayana Penukonda}, \bibinfo{person}{Andy Phelps},
  \bibinfo{person}{Jonathan Ross}, \bibinfo{person}{Matt Ross},
  \bibinfo{person}{Amir Salek}, \bibinfo{person}{Emad Samadiani},
  \bibinfo{person}{Chris Severn}, \bibinfo{person}{Gregory Sizikov},
  \bibinfo{person}{Matthew Snelham}, \bibinfo{person}{Jed Souter},
  \bibinfo{person}{Dan Steinberg}, \bibinfo{person}{Andy Swing},
  \bibinfo{person}{Mercedes Tan}, \bibinfo{person}{Gregory Thorson},
  \bibinfo{person}{Bo Tian}, \bibinfo{person}{Horia Toma},
  \bibinfo{person}{Erick Tuttle}, \bibinfo{person}{Vijay Vasudevan},
  \bibinfo{person}{Richard Walter}, \bibinfo{person}{Walter Wang},
  \bibinfo{person}{Eric Wilcox}, {and} \bibinfo{person}{Doe~Hyun Yoon}.}
  \bibinfo{year}{2017}\natexlab{}.
\newblock \showarticletitle{In-Datacenter Performance Analysis of a Tensor
  Processing Unit}. In \bibinfo{booktitle}{\emph{Proceedings of the 44th Annual
  International Symposium on Computer Architecture, {ISCA} 2017, Toronto, ON,
  Canada, June 24-28, 2017}}. \bibinfo{publisher}{{ACM}},
  \bibinfo{pages}{1--12}.
\newblock
\urldef\tempurl%
\url{https://doi.org/10.1145/3079856.3080246}
\showDOI{\tempurl}


\bibitem[Kandemir et~al\mbox{.}(1999)]%
        {kandemir1999linear}
\bibfield{author}{\bibinfo{person}{Mahmut Kandemir}, \bibinfo{person}{Alok
  Choudhary}, \bibinfo{person}{Nagaraj Shenoy}, \bibinfo{person}{Prithviraj
  Banerjee}, {and} \bibinfo{person}{J Ramenujarn}.}
  \bibinfo{year}{1999}\natexlab{}.
\newblock \showarticletitle{A linear algebra framework for automatic
  determination of optimal data layouts}.
\newblock \bibinfo{journal}{\emph{IEEE Transactions on Parallel and Distributed
  Systems}} \bibinfo{volume}{10}, \bibinfo{number}{2} (\bibinfo{year}{1999}),
  \bibinfo{pages}{115--135}.
\newblock


\bibitem[Khamis et~al\mbox{.}(2020)]%
        {khamis2020learning}
\bibfield{author}{\bibinfo{person}{Mahmoud~Abo Khamis}, \bibinfo{person}{Hung~Q
  Ngo}, \bibinfo{person}{XuanLong Nguyen}, \bibinfo{person}{Dan Olteanu}, {and}
  \bibinfo{person}{Maximilian Schleich}.} \bibinfo{year}{2020}\natexlab{}.
\newblock \showarticletitle{Learning models over relational data using sparse
  tensors and functional dependencies}.
\newblock \bibinfo{journal}{\emph{ACM Transactions on Database Systems (TODS)}}
  \bibinfo{volume}{45}, \bibinfo{number}{2} (\bibinfo{year}{2020}),
  \bibinfo{pages}{1--66}.
\newblock


\bibitem[Khamis et~al\mbox{.}(2016)]%
        {DBLP:conf/pods/KhamisNR16}
\bibfield{author}{\bibinfo{person}{Mahmoud~Abo Khamis},
  \bibinfo{person}{Hung~Q. Ngo}, {and} \bibinfo{person}{Atri Rudra}.}
  \bibinfo{year}{2016}\natexlab{}.
\newblock \showarticletitle{{FAQ:} Questions Asked Frequently}. In
  \bibinfo{booktitle}{\emph{Proceedings of the 35th {ACM} {SIGMOD-SIGACT-SIGAI}
  Symposium on Principles of Database Systems, {PODS} 2016, San Francisco, CA,
  USA, June 26 - July 01, 2016}}, \bibfield{editor}{\bibinfo{person}{Tova Milo}
  {and} \bibinfo{person}{Wang{-}Chiew Tan}} (Eds.). \bibinfo{publisher}{{ACM}},
  \bibinfo{pages}{13--28}.
\newblock
\urldef\tempurl%
\url{https://doi.org/10.1145/2902251.2902280}
\showDOI{\tempurl}


\bibitem[Kjolstad et~al\mbox{.}(2017)]%
        {kjolstad:2017:taco}
\bibfield{author}{\bibinfo{person}{Fredrik Kjolstad}, \bibinfo{person}{Shoaib
  Kamil}, \bibinfo{person}{Stephen Chou}, \bibinfo{person}{David Lugato}, {and}
  \bibinfo{person}{Saman Amarasinghe}.} \bibinfo{year}{2017}\natexlab{}.
\newblock \showarticletitle{The Tensor Algebra Compiler}.
\newblock \bibinfo{journal}{\emph{Proc. ACM Program. Lang.}}
  \bibinfo{volume}{1}, \bibinfo{number}{OOPSLA}, Article
  \bibinfo{articleno}{77} (\bibinfo{date}{Oct.} \bibinfo{year}{2017}),
  \bibinfo{numpages}{29}~pages.
\newblock
\showISSN{2475-1421}
\urldef\tempurl%
\url{https://doi.org/10.1145/3133901}
\showDOI{\tempurl}


\bibitem[Mart{\'\i}n-Garc{\'\i}a(2008)]%
        {martin2008xperm}
\bibfield{author}{\bibinfo{person}{Jos{\'e}~M Mart{\'\i}n-Garc{\'\i}a}.}
  \bibinfo{year}{2008}\natexlab{}.
\newblock \showarticletitle{xPerm: fast index canonicalization for tensor
  computer algebra}.
\newblock \bibinfo{journal}{\emph{Computer physics communications}}
  \bibinfo{volume}{179}, \bibinfo{number}{8} (\bibinfo{year}{2008}),
  \bibinfo{pages}{597--603}.
\newblock


\bibitem[Nikolic and Olteanu(2018)]%
        {10.1145/3183713.3183758}
\bibfield{author}{\bibinfo{person}{Milos Nikolic} {and} \bibinfo{person}{Dan
  Olteanu}.} \bibinfo{year}{2018}\natexlab{}.
\newblock \showarticletitle{Incremental View Maintenance with Triple Lock
  Factorization Benefits}. In \bibinfo{booktitle}{\emph{Proceedings of the 2018
  International Conference on Management of Data}} (Houston, TX, USA)
  \emph{(\bibinfo{series}{SIGMOD '18})}. \bibinfo{publisher}{Association for
  Computing Machinery}, \bibinfo{address}{New York, NY, USA},
  \bibinfo{pages}{365–380}.
\newblock
\showISBNx{9781450347037}
\urldef\tempurl%
\url{https://doi.org/10.1145/3183713.3183758}
\showDOI{\tempurl}


\bibitem[Paszke et~al\mbox{.}(2019)]%
        {Paszke_PyTorch_An_Imperative_2019}
\bibfield{author}{\bibinfo{person}{Adam Paszke}, \bibinfo{person}{Sam Gross},
  \bibinfo{person}{Francisco Massa}, \bibinfo{person}{Adam Lerer},
  \bibinfo{person}{James Bradbury}, \bibinfo{person}{Gregory Chanan},
  \bibinfo{person}{Trevor Killeen}, \bibinfo{person}{Zeming Lin},
  \bibinfo{person}{Natalia Gimelshein}, \bibinfo{person}{Luca Antiga},
  \bibinfo{person}{Alban Desmaison}, \bibinfo{person}{Andreas Kopf},
  \bibinfo{person}{Edward Yang}, \bibinfo{person}{Zachary DeVito},
  \bibinfo{person}{Martin Raison}, \bibinfo{person}{Alykhan Tejani},
  \bibinfo{person}{Sasank Chilamkurthy}, \bibinfo{person}{Benoit Steiner},
  \bibinfo{person}{Lu Fang}, \bibinfo{person}{Junjie Bai}, {and}
  \bibinfo{person}{Soumith Chintala}.} \bibinfo{year}{2019}\natexlab{}.
\newblock \showarticletitle{{PyTorch: An Imperative Style, High-Performance
  Deep Learning Library}}. In \bibinfo{booktitle}{\emph{Advances in Neural
  Information Processing Systems 32}}. \bibinfo{publisher}{Curran Associates,
  Inc.}, \bibinfo{pages}{8024--8035}.
\newblock
\urldef\tempurl%
\url{http://papers.neurips.cc/paper/9015-pytorch-an-imperative-style-high-performance-deep-learning-library.pdf}
\showURL{%
\tempurl}


\bibitem[Ran et~al\mbox{.}(2020)]%
        {ran2020tensor}
\bibfield{author}{\bibinfo{person}{Shi-Ju Ran}, \bibinfo{person}{Emanuele
  Tirrito}, \bibinfo{person}{Cheng Peng}, \bibinfo{person}{Xi Chen},
  \bibinfo{person}{Luca Tagliacozzo}, \bibinfo{person}{Gang Su}, {and}
  \bibinfo{person}{Maciej Lewenstein}.} \bibinfo{year}{2020}\natexlab{}.
\newblock \bibinfo{booktitle}{\emph{Tensor network contractions: methods and
  applications to quantum many-body systems}}.
\newblock \bibinfo{publisher}{Springer Nature}.
\newblock


\bibitem[Schleich et~al\mbox{.}(2022)]%
        {schleich2022optimizing}
\bibfield{author}{\bibinfo{person}{Maximilian Schleich}, \bibinfo{person}{Amir
  Shaikhha}, {and} \bibinfo{person}{Dan Suciu}.}
  \bibinfo{year}{2022}\natexlab{}.
\newblock \showarticletitle{Optimizing Tensor Programs on Flexible Storage}.
\newblock \bibinfo{journal}{\emph{arXiv preprint arXiv:2210.06267}}
  (\bibinfo{year}{2022}).
\newblock


\bibitem[Shaikhha et~al\mbox{.}(2022)]%
        {shaikhha2022functional}
\bibfield{author}{\bibinfo{person}{Amir Shaikhha}, \bibinfo{person}{Mathieu
  Huot}, \bibinfo{person}{Jaclyn Smith}, {and} \bibinfo{person}{Dan Olteanu}.}
  \bibinfo{year}{2022}\natexlab{}.
\newblock \showarticletitle{Functional collection programming with semi-ring
  dictionaries}.
\newblock \bibinfo{journal}{\emph{Proceedings of the ACM on Programming
  Languages}} \bibinfo{volume}{6}, \bibinfo{number}{OOPSLA1}
  (\bibinfo{year}{2022}), \bibinfo{pages}{1--33}.
\newblock


\bibitem[Smith and Gray(2018)]%
        {DBLP:journals/jossw/SmithG18}
\bibfield{author}{\bibinfo{person}{Daniel G.~A. Smith} {and}
  \bibinfo{person}{Johnnie Gray}.} \bibinfo{year}{2018}\natexlab{}.
\newblock \showarticletitle{opt{\_}einsum - {A} Python package for optimizing
  contraction order for einsum-like expressions}.
\newblock \bibinfo{journal}{\emph{J. Open Source Softw.}} \bibinfo{volume}{3},
  \bibinfo{number}{26} (\bibinfo{year}{2018}), \bibinfo{pages}{753}.
\newblock
\urldef\tempurl%
\url{https://doi.org/10.21105/joss.00753}
\showDOI{\tempurl}


\bibitem[Spampinato and P{\"{u}}schel(2016)]%
        {spampinato2016basic}
\bibfield{author}{\bibinfo{person}{Daniele~G. Spampinato} {and}
  \bibinfo{person}{Markus P{\"{u}}schel}.} \bibinfo{year}{2016}\natexlab{}.
\newblock \showarticletitle{A basic linear algebra compiler for structured
  matrices}. In \bibinfo{booktitle}{\emph{Proceedings of the 2016 International
  Symposium on Code Generation and Optimization, {CGO} 2016, Barcelona, Spain,
  March 12-18, 2016}}, \bibfield{editor}{\bibinfo{person}{Bj{\"{o}}rn Franke},
  \bibinfo{person}{Youfeng Wu}, {and} \bibinfo{person}{Fabrice Rastello}}
  (Eds.). \bibinfo{publisher}{{ACM}}, \bibinfo{pages}{117--127}.
\newblock
\urldef\tempurl%
\url{https://doi.org/10.1145/2854038.2854060}
\showDOI{\tempurl}


\bibitem[Strout et~al\mbox{.}(2018)]%
        {strout2018sparse}
\bibfield{author}{\bibinfo{person}{Michelle~Mills Strout},
  \bibinfo{person}{Mary Hall}, {and} \bibinfo{person}{Catherine Olschanowsky}.}
  \bibinfo{year}{2018}\natexlab{}.
\newblock \showarticletitle{The sparse polyhedral framework: Composing
  compiler-generated inspector-executor code}.
\newblock \bibinfo{journal}{\emph{Proc. IEEE}} \bibinfo{volume}{106},
  \bibinfo{number}{11} (\bibinfo{year}{2018}), \bibinfo{pages}{1921--1934}.
\newblock


\bibitem[Tang et~al\mbox{.}(2020)]%
        {DBLP:journals/cgf/TangSKPLP20}
\bibfield{author}{\bibinfo{person}{Xuan Tang}, \bibinfo{person}{Teseo
  Schneider}, \bibinfo{person}{Shoaib Kamil}, \bibinfo{person}{Aurojit Panda},
  \bibinfo{person}{Jinyang Li}, {and} \bibinfo{person}{Daniele Panozzo}.}
  \bibinfo{year}{2020}\natexlab{}.
\newblock \showarticletitle{{EGGS:} Sparsity-Specific Code Generation}.
\newblock \bibinfo{journal}{\emph{Comput. Graph. Forum}} \bibinfo{volume}{39},
  \bibinfo{number}{5} (\bibinfo{year}{2020}), \bibinfo{pages}{209--219}.
\newblock
\urldef\tempurl%
\url{https://doi.org/10.1111/cgf.14080}
\showDOI{\tempurl}


\bibitem[Titov et~al\mbox{.}(2013)]%
        {titov2013generating}
\bibfield{author}{\bibinfo{person}{Alexey~V Titov}, \bibinfo{person}{Ivan~S
  Ufimtsev}, \bibinfo{person}{Nathan Luehr}, {and} \bibinfo{person}{Todd~J
  Martinez}.} \bibinfo{year}{2013}\natexlab{}.
\newblock \showarticletitle{Generating efficient quantum chemistry codes for
  novel architectures}.
\newblock \bibinfo{journal}{\emph{Journal of chemical theory and computation}}
  \bibinfo{volume}{9}, \bibinfo{number}{1} (\bibinfo{year}{2013}),
  \bibinfo{pages}{213--221}.
\newblock


\bibitem[Vasilache et~al\mbox{.}(2018)]%
        {vasilache2018tensor}
\bibfield{author}{\bibinfo{person}{Nicolas Vasilache},
  \bibinfo{person}{Oleksandr Zinenko}, \bibinfo{person}{Theodoros Theodoridis},
  \bibinfo{person}{Priya Goyal}, \bibinfo{person}{Zachary DeVito},
  \bibinfo{person}{William~S Moses}, \bibinfo{person}{Sven Verdoolaege},
  \bibinfo{person}{Andrew Adams}, {and} \bibinfo{person}{Albert Cohen}.}
  \bibinfo{year}{2018}\natexlab{}.
\newblock \showarticletitle{Tensor comprehensions: Framework-agnostic
  high-performance machine learning abstractions}.
\newblock \bibinfo{journal}{\emph{arXiv preprint arXiv:1802.04730}}
  (\bibinfo{year}{2018}).
\newblock


\bibitem[Verdoolaege(2010)]%
        {VERDOOLAEGE2010ISL}
\bibfield{author}{\bibinfo{person}{Sven Verdoolaege}.}
  \bibinfo{year}{2010}\natexlab{}.
\newblock \showarticletitle{isl: An Integer Set Library for the Polyhedral
  Model}. In \bibinfo{booktitle}{\emph{Mathematical Software (ICMS'10)}}
  \emph{(\bibinfo{series}{LNCS 6327})},
  \bibfield{editor}{\bibinfo{person}{Komei Fukuda}, \bibinfo{person}{Joris
  Hoeven}, \bibinfo{person}{Michael Joswig}, {and} \bibinfo{person}{Nobuki
  Takayama}} (Eds.). \bibinfo{publisher}{Springer-Verlag},
  \bibinfo{pages}{299--302}.
\newblock


\end{thebibliography}
